\documentclass[reprint,superscriptaddress,onecolumn]{revtex4-2}

\usepackage[]{graphicx} 
\usepackage[english]{babel}
\usepackage{amsmath}
\usepackage{amssymb}
\usepackage[colorlinks=true, allcolors=blue]{hyperref}
\usepackage{cleveref}
\usepackage{siunitx}

\usepackage[table]{xcolor}
\usepackage{makecell}
\usepackage{multirow}   

\usepackage[normalem]{ulem}

\newcommand{\lp}{\left(}
\newcommand{\rp}{\right)}
\newcommand{\lbk}{\left[}
\newcommand{\rbk}{\right]}

\begin{document}

\title{Cell size control in bacteria is modulated through extrinsic noise, single-cell- and population-growth}
\author{Arthur Genthon}
\affiliation{Max Planck Institute for the Physics of Complex Systems, 01187 Dresden, Germany}
\author{Philipp Thomas}
\affiliation{Department of Mathematics, Imperial College London, London, United Kingdom}

\begin{abstract}
    Living cells maintain size homeostasis by actively compensating for size fluctuations. 
    Here, we present two stochastic maps that unify phenomenological models by integrating fluctuating single-cell growth rates and size-dependent noise mechanisms with cell size control. One map is applicable to mother machine lineages and the other to lineage trees of exponentially-growing cell populations, which reveals that population dynamics alter size control measured in mother machine experiments.
    For example, an adder can become more sizer-like or more timer-like at the population level depending on the noise statistics.
    Our analysis of bacterial data identifies extrinsic noise as the dominant mechanism of size variability, characterized by a quadratic conditional variance-mean relationship for division size across growth conditions. This finding contradicts the reported independence of added size relative to birth size but is consistent with the adder property in terms of the independence of the mean added size. 
    Finally, we derive a trade-off between population-growth-rate gain and division-size noise. 
    Correlations between size control quantifiers and single-cell growth rates inferred from data indicate that bacteria prioritize a narrow division-size distribution over growth rate maximisation.
\end{abstract}

\maketitle

\section{Introduction}

Cell-size homeostasis is the process by which cells maintain a stable size distribution over successive generations \cite{cadart_physics_2019}. In bacteria, this distribution is relatively narrow, implying that differences in size are actively and gradually corrected \cite{chien_cell_2012}. The development of specialized microfluidic devices has enabled long-term single-cell measurements of growth and division, providing direct access to single-cell variability and offering an opportunity to elucidate the underlying mechanisms \cite{wang_robust_2010,hashimoto_noise-driven_2016}. 

A hallmark of cell size control is the discovery of the adder mechanism \cite{taheri-araghi_cell-size_2015,campos_constant_2014}, whereby cells add, on average, a fixed length between birth and division, largely independent of their length at birth. Adder behaviour has been observed across a wide range of organisms \cite{sauls_adder_2016}, including bacteria \cite{campos_constant_2014,taheri-araghi_cell-size_2015,deforet_cell-size_2015,tanouchi_noisy_2015}, mycobacteria \cite{priestman_mycobacteria_2017}, budding yeast \cite{soifer_single-cell_2016}, archaea \cite{eun_archaeal_2017}, and mammalian cells \cite{cadart_size_2018}. 
A mechanistic interpretation of the adder is that division is triggered when a division-related protein accumulates to a threshold \cite{deforet_cell-size_2015,ghusinga_mechanistic_2016,si_mechanistic_2019}. However, standard experimental validations of the adder and sizer mechanisms rely on conditional means or on linear regression of cell size alone \cite{tanouchi_noisy_2015,priestman_mycobacteria_2017}, and it remains unclear to what extent the full distributions of single-cell observables are consistent with simple cell-size control models.
Motivated by the adder phenomenology, several stochastic models of cell-size control have been proposed. A widely used framework is the noisy linear map \cite{tanouchi_noisy_2015}, a simple auto-regressive process in which the division size is expressed as a linear function of the birth size, with a memory-less noise term added independently of cell size \cite{tanouchi_noisy_2015}. The predictions of this phenomenological model are also consistent with the threshold-crossing model \cite{ghusinga_mechanistic_2016}. 
Other approaches, including time-additive noise models, can also reproduce adder-like behaviour at the level of conditional means \cite{amir_cell_2014}. 
While these models can all account for average trends, it is unclear to what extent they capture higher-order statistics of the data. 

This discrepancy is highlighted by two foundational studies of the adder mechanism. \emph{Taheri et al.} reported that the added size between birth and division is independent of birth size, suggesting that size increments themselves are the primary controlled variable \cite{taheri-araghi_cell-size_2015}. On the other hand, \emph{Amir} introduced a time-additive noise model in which stochasticity acts on the duration of the cell cycle rather than directly on size \cite{amir_cell_2014}. Although this model reproduces adder-like behaviour at the level of conditional means and yields realistic size distributions, it is not immediately compatible with the reported collapse of added-size distributions. Crucially, despite their different interpretations, both approaches effectively treat cell-size control as decoupled from other physiological variables. In particular, they neglect fluctuations in single-cell growth rates, which modulate cell size and induce correlations \cite{modi_analysis_2017,grilli_empirical_2018,genthon_noisy_2025}. The apparent agreement at the average level could mask fundamentally different underlying mechanisms, underscoring the need for models that explicitly couple size control with growth rate.

\begin{figure}
    \centering
    \includegraphics[width=0.9\linewidth]{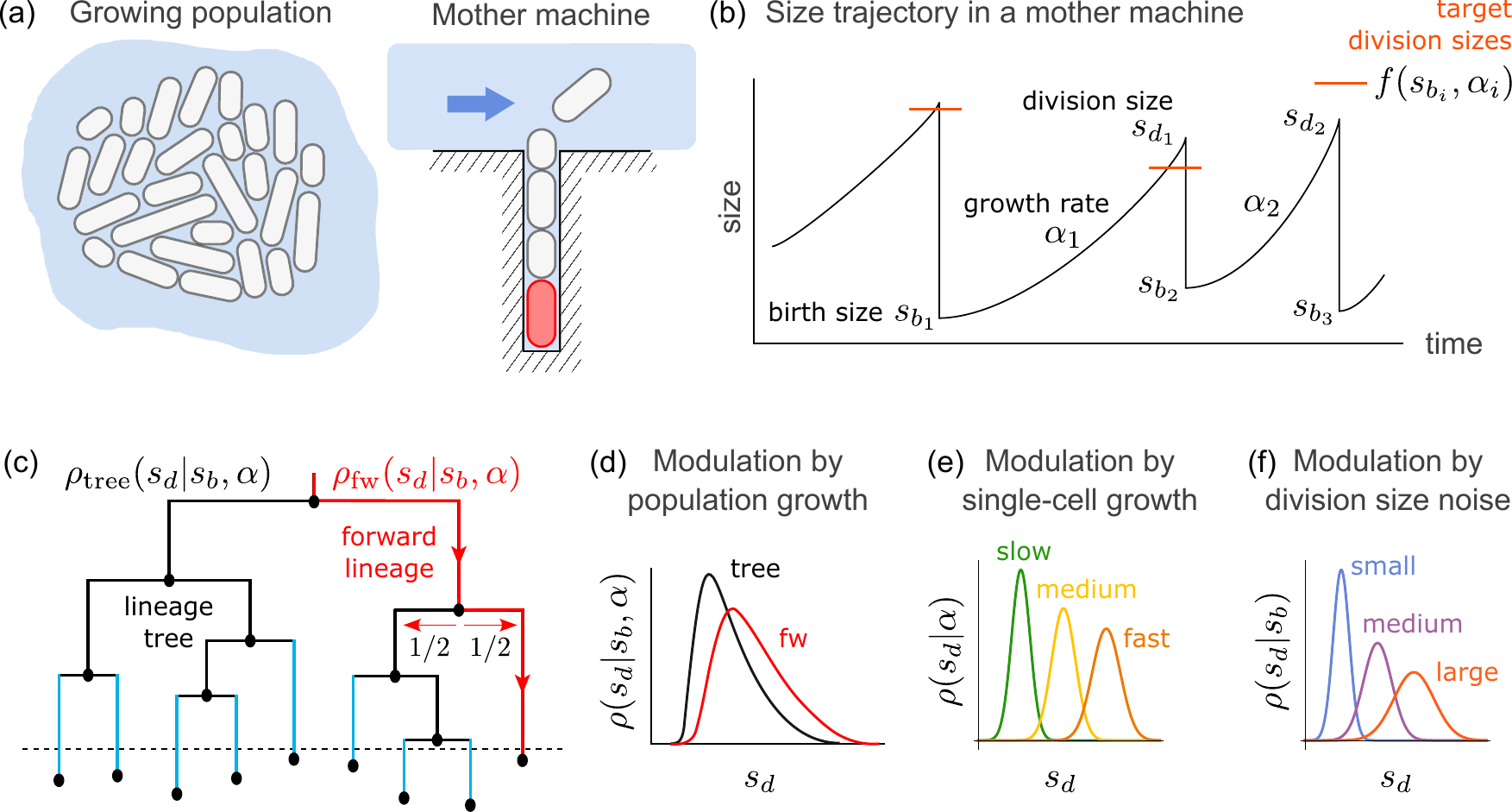}
    \caption{\textbf{Modulation of cell size control by noise, growth rate and experimental setup.} (a) Experimental setups: growing population and mother machine. (b) Cell size trajectory along a single lineage in a mother machine. (c) The tree statistics is obtained by sampling with uniform weights all cell divisions in the population, shown in black, but excluding leaf cells, shown in light blue, that have not yet divided at the end of the experiment (dotted line). The forward statistics can be obtained in mother machines or with a non-uniform sampling of lineages in the population \cite{nozoe_inferring_2017}. (d) The tree and forward conditional distributions of division sizes are different. (e,f) Cell size control can be modulated at the average and variance levels by (e) single cell growth rates $\alpha$ and (f) noise in division sizes. }
    \label{fig:intro_panel}
\end{figure}

Another issue is the distinction between lineage-based and population-based quantification of cell size control, 
which remains unexplored territory. 
Many theoretical models implicitly assume that statistics measured along a single lineage, such as those obtained in mother-machine experiments, are equivalent to statistics measured in an exponentially growing population (\cref{fig:intro_panel}a-b). However, it is well established that growing populations are subject to selection effects, with faster-growing cells being overrepresented (\cref{fig:intro_panel}c). This leads to systematic biases between lineage and population statistics for observables, including generation times \cite{powell_growth_1956,hashimoto_noise-driven_2016,sughiyama_fitness_2019,genthon_fluctuation_2020}, birth and division sizes \cite{thomas_single-cell_2017,thomas_analysis_2018}, snapshot sizes \cite{genthon_analytical_2022}, division kernels \cite{thomas_single-cell_2017,thomas_analysis_2018,sughiyama_fitness_2019} and single-cell growth rates \cite{priestman_mycobacteria_2017,thomas_single-cell_2017} (\cref{fig:intro_panel}d). 
Despite this, the consequences of lineage-population bias for inferring cell-size control strategies have received comparatively little attention.
These considerations raise fundamental questions of which sources of noise dominate cell size control and how cell size control can be interpreted in the presence of growth-rate variability and population-level selection. 

In this article, we address these questions by generalising the noisy linear map to quantify the effects of competing noise mechanisms  of bacterial growth in mother machine lineages, and we accompany this model with a corresponding tilted linear map that holds in lineage trees. We show that a specific scaling of division size fluctuations leads to a characteristic collapse of the conditional variance onto a single curve, consistent with available experimental data. We further demonstrate that single-cell growth rate and population growth introduce a bias in the inferred relationship between birth and division sizes, thereby altering the apparent cell size control mechanism. This implies that size-control strategies extracted from lineage trees and from single-lineage measurements are not directly comparable. Our results highlight the interplay between stochastic size control, growth rate, and population dynamics (\cref{fig:intro_panel}d-f), and provide analytical insights and new interpretations of the mechanisms governing cell-size homeostasis using single-cell data.

\section{Results}

\subsection{Generalised noisy linear map of cell size control}

Many evolutionarily divergent organisms employ similar mechanisms of cell size control \cite{sauls_adder_2016}. These mechanisms are quantitatively captured by a map relating the cell size at birth $s_b$ to the target size at division \cite{amir_cell_2014,tanouchi_noisy_2015} (\cref{fig:intro_panel}b). We postulate that the average map depends linearly on $s_b$ to include the common adder and sizer mechanisms \cite{tanouchi_noisy_2015} and also depends on the single-cell growth rate $\alpha$  (\cref{fig:intro_panel}e) to model its dependence on the energy status of a cell \cite{modi_analysis_2017,grilli_empirical_2018,genthon_noisy_2025}.
The general model of cell size control is then:
\begin{subequations}
\label{eq_csc_model}
\begin{align}
\label{eq_lin_map}
s_d &= f(s_b,\alpha) +\eta(s_b,\alpha) \\
f(s_b,\alpha) &= a(\alpha) s_b + b(\alpha) \\
\eta(s_b,\alpha) &= \eta_a(\alpha) + \sqrt{f(s_b,\alpha)} \eta_i(\alpha) + f(s_b,\alpha) \eta_e(\alpha) \,.
\label{eq_noise_ampl}
\end{align}
\end{subequations}
where the slope $a$ defines the mechanism of cell size control (sizer $a=0$, adder $a=1$, timer $a=2$).

Crucially, our generalised framework includes several size-dependent noise sources (\cref{fig:intro_panel}f) in addition to the commonly considered additive noise \cite{tanouchi_noisy_2015,modi_analysis_2017,priestman_mycobacteria_2017,thomas_analysis_2018,genthon_noisy_2025}. Fluctuations in division sizes $s_d$ are modelled with a general combination of zero-mean and independent additive noise $\eta_a(\alpha)$, intrinsic noise $\eta_i(\alpha)$ and extrinsic noise $\eta_e(\alpha)$, with the scaling $\mathrm{CV}^2[s_d|s_b,\alpha]=\sigma_a^2(\alpha)/f(s_b,\alpha)^2+\sigma_i^2(\alpha)/f(s_b,\alpha)+\sigma_e^2(\alpha)$ where $\sigma_k^2$ is the variance of noise $\eta_k$ for $k \in \{a,i,e\}$, illustrated in \cref{fig:pop_alpha_cst}a-c. Note that just like the target division size $f(s_b,\alpha)$, the noise terms can be modulated by single-cell growth rates $\alpha$. Since noise has a zero mean, the target division size and the mean division size $\langle s_d | s_b,\alpha \rangle$, indicated by angle brackets, are used interchangeably in the following.

Our noise model can be viewed as a molecular regulator setting the division size. It is common in the stochastic gene expression literature to break down variability into intrinsic noise arising from fluctuations in molecular circuit dynamics and extrinsic noise from upstream factors and cell-specific factors \cite{elowitz_stochastic_2002,paulsson_summing_2004,hilfinger_separating_2011,thomas_intrinsic_2019}. This decomposition typically leads to relative noise amplitudes for the stochastic variable (measured by the coefficient of variation CV=std/mean) scaling as the inverse of the square root of the variable mean for intrinsic noise, and independent of the mean for extrinsic noise. Assuming constant concentrations of the regulator, these scalings translate into the noise amplitudes in \cref{eq_noise_ampl}, and we include the customary additive noise term for completeness.

\subsection{Tilted linear map predicts that noise changes the mode of cell size control in populations}
\label{sec_lin_pop_bias}

\begin{figure*}
    \includegraphics[width=\linewidth]{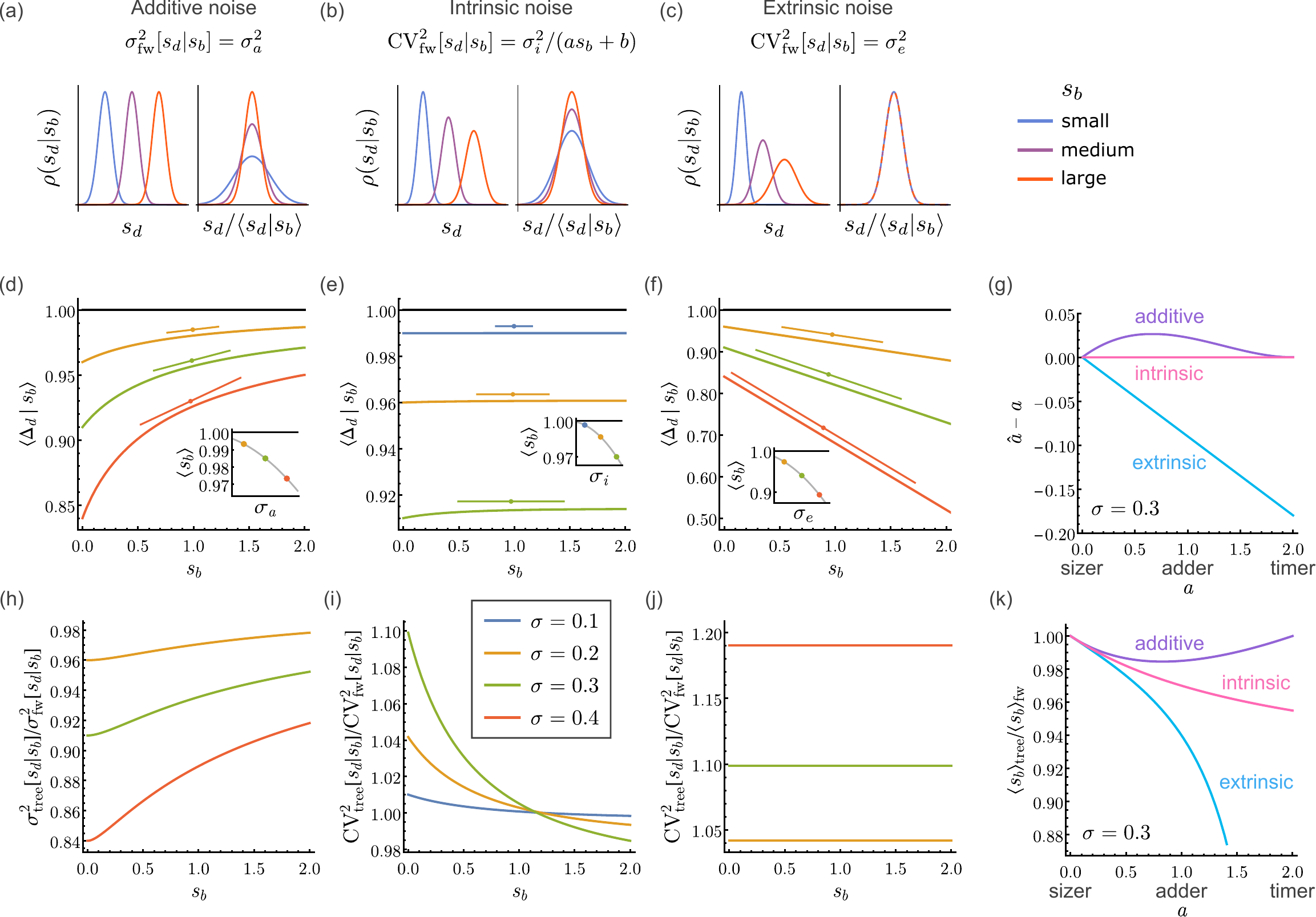}
    \caption{\textbf{Noise changes the mode of cell size control in populations.} In each of the first three columns only one type of noise is considered: additive, intrinsic or extrinsic. 
    (a-c) Different types of noise impact differently the mean and variance of conditional distributions of division sizes.
    (d-f) The conditional average added size in the tree statistics depends on the size at birth and on the noise levels (boxed legend) for a reference forward adder ($a=1$, black). 
    Predictions of the population linear map (slope $\hat{a}$ from \cref{eq_pop_lin_a}, offset for readability) shown for sizes in an interval of two standard deviations below and above the mean birth size (circles) agree with exact solutions computed with \cref{powell_cond_sd}.
    Insets show average birth sizes in the tree statistics against noise levels, compared to the forward value (black). 
    (h-j) Modulation of the population variance of division sizes. Variance (h) and CV (i,j) on the y-axis are rescaled by their values in the forward statistics to highlight the modulation of the birth-size dependence. Extrinsic noise remains extrinsic while extra size-dependence arise for forward additive and intrinsic noises.
    (g,k) The last column illustrates the forward-to-population changes in (g) linear map slope $\hat{a}-a$ and in (k) mean birth size $\langle s_b \rangle_\mathrm{tree}/\langle s_b \rangle_\mathrm{fw}$, as a function of the forward slope $a$ and for the three types of noise with the same standard deviation $\sigma=0.3$.}
    \label{fig:pop_alpha_cst}
\end{figure*}

It is well understood that single-lineage statistics in mother machines differ from those of lineages in growing populations, in which cells with above-average reproductive success are overrepresented. This bias has been studied for many observables \cite{powell_growth_1956,hashimoto_noise-driven_2016,sughiyama_fitness_2019,genthon_fluctuation_2020,thomas_single-cell_2017,thomas_analysis_2018,genthon_analytical_2022,sughiyama_fitness_2019}, and we show here that the mode of cell size control itself is biased in tree statistics compared to forward lineages (see SM and \cref{powell_cond_sd} in the Methods section), as illustrated in \cref{fig:intro_panel}d.
When cell size is tightly regulated \cite{taheri-araghi_cell-size_2015}, a small noise analysis reveals that cell size control at the population level also follows a noisy linear map, which is tilted compared to the linear map along forward lineages \cref{eq_csc_model}. 
In the following, we consider single-cell exponential growth, which characterizes a broad range of organisms \cite{taheri-araghi_cell-size_2015,priestman_mycobacteria_2017,soifer_single-cell_2016,eun_archaeal_2017}, with no fluctuations in single-cell growth rates (this assumption will be relaxed in \cref{sec_grow_rate}). In this case, the tilted linear map takes the form $ s_d = \hat{a} s_b + \hat{b} + \hat{\eta}$, with modified coefficients and noise given at leading order in the noise variances by (see Methods):
\begin{subequations}
\label{eq_pop_lin_map}
    \begin{align}
    \label{eq_pop_lin_a}
    \hat{a} &\approx a \lbk 1 + \frac{(2-a)^2}{4b^2} \sigma_a^2 - \sigma_e^2 \rbk - \frac{(2-a)a^2}{(2+a)b^2}   \sigma_a^2   \mathrm{CV}_p^2
     \\
     \label{eq_pop_lin_b}
    \hat{b} &\approx b \lbk 1 - \frac{4-a^2}{4b^2} \sigma_a^2 -\frac{1}{b}\sigma_i^2 - \sigma_e^2  \rbk +\frac{a^2}{(2+a)b}\sigma_a^2  \mathrm{CV}_p^2 \\
    \label{eq_pop_lin_eta}
    \langle \hat{\eta}^2|s_b\rangle_{\mathrm{tree}} &\approx \lp 1 -\frac{\gamma_a \sigma_a}{a s_b +b} \rp \sigma_a^2 + (a s_b+b) \lp 1 -\frac{ \gamma_i \sigma_i}{\sqrt{a s_b +b}} \rp \sigma_i^2 + (a s_b+b)^2 \lp 1 -\gamma_e \sigma_e \rp \sigma_e^2\,,
\end{align}
\end{subequations}
where $\mathrm{CV}_p$ is the coefficient of variation of the septum position (asymmetry in birth sizes between the two daughter cells), and $\gamma_k$ is the skewness of noise $\eta_k$.
Let us explore the main consequences of this tilted linear map. 

First, the mechanism of cell size control in the tree statistics, $\hat{a}$, is modulated by noise on division size, as shown in \cref{fig:pop_alpha_cst}g. Additive and extrinsic noises have opposite effects, respectively increasing and decreasing $\hat{a}$ compared to its forward counterpart $a$; while intrinsic noise has a negligible effect (see SM for next order correction in $\sigma_i^3$). The modulation of size control is more significant for weaker size controls (increasing $a$) for the extrinsic noise model. On the other hand, the modulation due to additive noise is maximal at intermediate $a=2/3$, thereby pushing the population mechanism towards the adder, and cancelling for the timer at $a=2$. Moreover, independently of the type of noise, a forward sizer remains a sizer at the population level. This is a direct consequence of \cref{powell_cond_sd}, valid more generally for a broad class of single-cell growth laws (see SM). 
Unlike the mechanism $a$, the intercept $b$ is reduced at the population level independently of the type of noise, and intrinsic noise has a non-vanishing effect. 

Second, slope and intercept modulations due to noise in division asymmetry vanish in the absence of additive noise and thus constitute next-order corrections (see SM for illustration). In particular, division asymmetry does not modulate size control at any order when coupled only to extrinsic noise. 

Third, the tilted linear map reveals how division size is biased for a given birth size, but birth size statistics are themselves modulated through population growth (see SM):
\begin{equation}
\label{eq_bias_sb}
    \langle s_b \rangle_{\mathrm{tree}} \approx \langle s_b \rangle_{\mathrm{fw}} \lbk  1 - \frac{a(2-a)}{2(2+a)b^2}\sigma_a^2 - \frac{a}{(2+a)b}\sigma_i^2 - \frac{2a}{4-a^2}\sigma_e^2 + \frac{4}{4-a^2} \mathrm{CV}_p^2 \rbk \,.
\end{equation}
Variability in division sizes decreases the mean birth size in the population statistics for all forms of noise, as shown in \cref{fig:pop_alpha_cst}k, while variability in septum position has the opposite effect. Therefore, the average birth size can be either smaller or larger in populations than in forward lineages, depending on noise levels. On the other hand, we prove in the SM that including the leaf cells in the tree statistics always decreases the average birth size. Like for size control, additive noise has a non-monotonic effect with a maximal modulation of the average birth size at $a=2(\sqrt{2}-1)$.

We illustrate the size control modulations of an adder in the forward statistics under different types of noise in \cref{fig:pop_alpha_cst}d-f.
The horizontal black lines indicate the independence of added size and birth size for the forward adder. The average added sizes in populations are computed by numerically solving \cref{powell_cond_sd} where $b+\eta_k$ is Gamma-distributed for different noises $k \in \{a,i,e\}$ and different noise levels indicated by colours. 
For all noise types and for all birth sizes, the expected added size is smaller in the population statistics than in forward lineages. The property $\langle \Delta_d |s_b \rangle_{\mathrm{tree}} \leq \langle \Delta_d |s_b \rangle_{\mathrm{fw}} $, observed here for a forward adder, is a general consequence of \cref{powell_cond_sd} valid independently of the size control mechanism. Please note that it does not imply an inequality for the marginal added sizes $\langle \Delta_d \rangle$ since the forward and population birth size distributions are different, as shown in \cref{eq_bias_sb}. Nevertheless, we derive an expression of $\langle \Delta_d \rangle_{\mathrm{tree}}$ in the SM and show that it is smaller than the forward average when $a \leq 1$, which includes the adder. 
The coloured straight lines show the modulations of the forward adder predicted by the tilted linear map \cref{eq_pop_lin_a} in the small noise regime, which are in excellent agreement with the exact solution. 
Note that the small discrepancy between the slow increase of mean added size with birth size observed for intrinsic noise with $\sigma_i=0.3$ and the independence predicted by the tilted linear map \cref{eq_pop_lin_a} [\cref{fig:pop_alpha_cst}e] is resolved with next-order corrections in $\sigma_i^3$ [SM].
The tilted linear predictions show intervals of two standard deviations around the tree average birth sizes indicated by circles (see SM). The average birth sizes are computed with \cref{eq_bias_sb} and are reported in the inset plots.

Finally, the type of noise on division size is also modulated at the population level (\cref{eq_pop_lin_eta}). A forward extrinsic noise remains extrinsic at the population level while
additive and intrinsic noises are not conserved at the population level, generating complex dependencies on birth size. Noise modulations are illustrated in \cref{fig:pop_alpha_cst}h-j.

\subsection{Extrinsic noise model captures a broad range of single-cell data}

\subsubsection{Bacterial growth data collapse across a broad range of conditions}

\begin{figure}
    \centering
    \includegraphics[width=\linewidth]{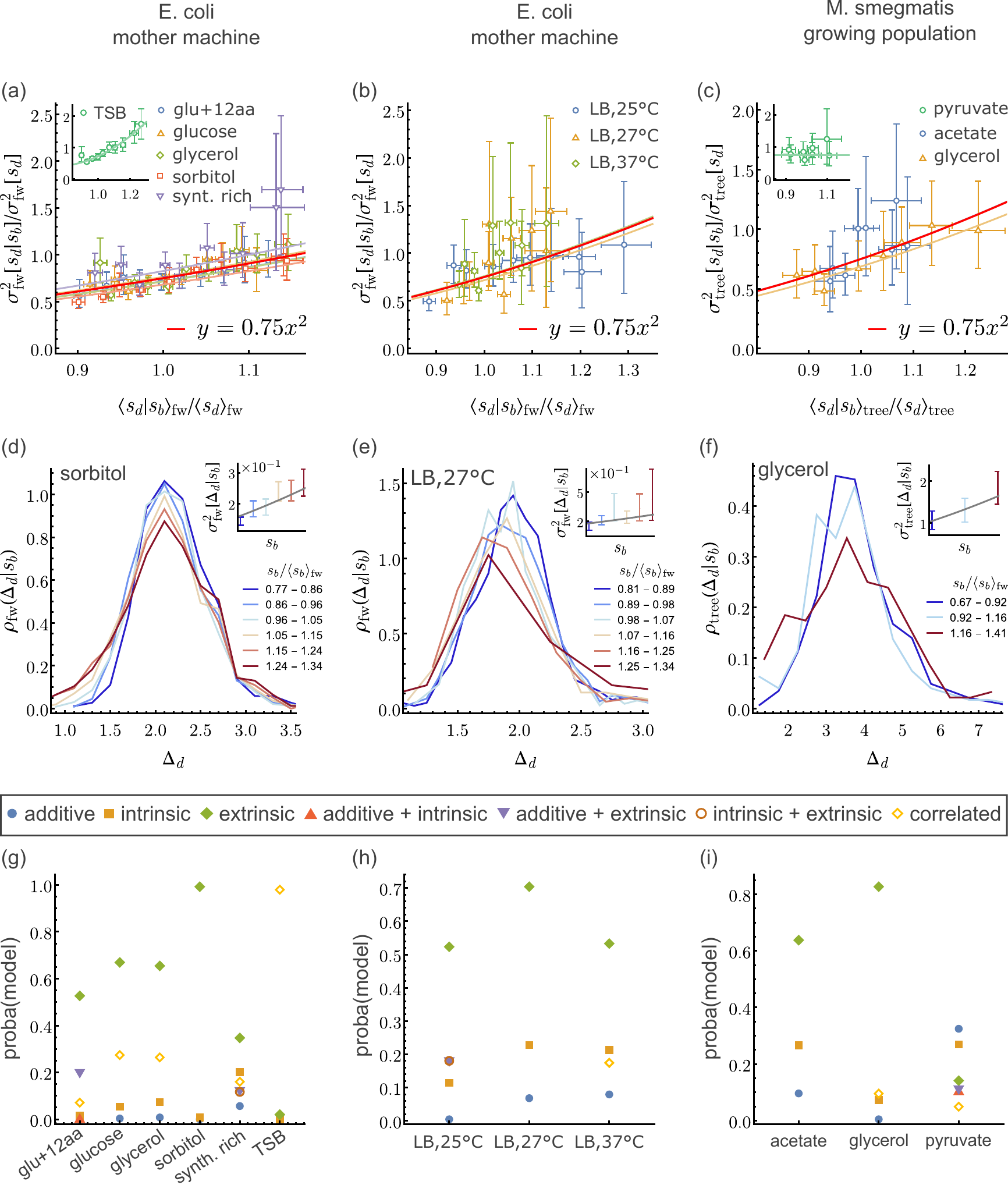}
    \caption{\textbf{Single-cell data of medium and temperature perturbations in \textit{E. coli} and \textit{M. smegmatis} collapse on extrinsic noise model predictions.} First column: data of \textit{E. coli} in mother machine in different media from \cite{taheri-araghi_cell-size_2015}, second column: data of \textit{E. coli} in mother machine in LB medium at different temperatures from \cite{tanouchi_noisy_2015}, third column: data of \textit{M. smegmatis} in growing population in different media from \cite{priestman_mycobacteria_2017}. (a-c) Collapse of conditional variance $\sigma^2[s_d|s_b]$ vs conditional mean $\langle s_d |s_b \rangle$, rescaled by marginal variance $\sigma^2[s_d]$ and mean $\langle s_d \rangle$, on extrinsic master curve $y=0.75 x^2$ for most conditions. The best extrinsic fits for each condition are shown as light-coloured curves, and the master curve is shown in red. Insets: conditions which do not follow the extrinsic noise model, with the same axis labels as the main plots, and the best noise model (TSB: correlated, pyruvate: additive) shown with light colour curves. Error bars denote the $95\%$ confidence interval of the estimate determined by bootstrapping. (d-f) Conditional distributions of added sizes for different birth sizes do not collapse. Inset: added size variance vs birth size follows the extrinsic model prediction shown in grey. (g-i) Likelihoods of different noise models were computed as Akaike weights $p(m)=\text{softmax}(-\mathrm{AIC}(m)/2)$.}
    \label{fig:data_noise}
\end{figure}

In the previous section, we demonstrated the importance of quantifying fluctuations in division sizes to understand modulations in cell size control and average birth size. 
Here, we quantify these fluctuations for mother machine data of \textit{Escherichia coli} in different growth media \cite{taheri-araghi_cell-size_2015} and in different temperatures \cite{tanouchi_noisy_2015}; and for data from growing populations of \textit{Mycobacterium smegmatis} in different growth media \cite{priestman_mycobacteria_2017}. 

To infer the form of the noise across conditions, we computed the conditional variance of division sizes $\sigma^2[s_d|s_b]$ and the conditional mean division size $\langle s_d|s_b \rangle$ for different birth sizes. The birth sizes were divided into $n$ uniform bins spanning the quantile range $[p,1-p]$ to exclude extreme-sized cells ($n=10$ for \textit{E. coli}, $n=8$ for \textit{M. smegmatis}, $p=0.025$ for Taheri's data, and $p=0.05$ for Tanouchi's and Priestman's data). Note that $\langle s_d|s_b \rangle$ was obtained from data and did not rely on any model of cell size control.
We then fitted $\sigma^2[s_d|s_b]$ against $\langle s_d|s_b \rangle$ with seven noise models: purely additive, purely intrinsic and purely extrinsic; independent combinations of additive and intrinsic, additive and extrinsic, intrinsic and extrinsic; and the most complete model with all three noises when also allowing for correlations between additive and extrinsic noises, which we call the correlated noise model. 
Model likelihoods were computed using the Akaike weights based on the Akaike information criterion (AIC), which is a trade-off between goodness of fit and model simplicity (number of parameters) \cite{burnham_multimodel_2004}.
We excluded redundant models that gave the same fit as a simpler model when one or more fitting parameters were found to be zero.

This analysis revealed that, for 10 of 13 conditions, the extrinsic noise model best describes the data. We excluded the condition glucose+6aa from \cite{taheri-araghi_cell-size_2015} because it did not exhibit any trend. The two other outliers are best described by a correlated noise model, for \textit{E. coli} in Tryptic Soy Broth (TSB), and by the additive noise model for \textit{M. smegmatis} in pyruvate. The model probabilities are reported in \cref{fig:data_noise}g-i, and fits and performances for individual conditions are shown in SM. 
We show in \cref{fig:data_noise}a-c that all data which are best described by the extrinsic noise model collapse on the master curve $\sigma^2[s_d|s_b]/\sigma^2[s_d] = 0.75 \langle s_d |s_b \rangle^2/\langle s_d \rangle^2$ when normalizing the axes by the marginal variance and mean. This collapse
is a prediction of the extrinsic noise model for the adder with negligible division asymmetry, in which case the extrinsic variance and the CV of division sizes are linked by the simple relation $\mathrm{CV}^2[s_d] = 4 \sigma_e^2 /3$ (see SM or \cref{eq_CV_sd}). The master curve is shown in red while the best fits of individual conditions are shown in light colours. 
The outlier conditions are shown in the insets in \cref{fig:data_noise}a,c.

Surprisingly, the extrinsic fluctuations observed above are incompatible with the collapse of added size distributions for different birth sizes proposed in the original adder paper \cite{taheri-araghi_cell-size_2015}. 
To clarify this issue, we reproduced the data analysis from \cite{taheri-araghi_cell-size_2015}. We show in \cref{fig:data_noise}d the distributions of added sizes for different birth sizes for \textit{E. coli} in sorbitol (see SM for other conditions), which indeed get wider as birth size increases. To quantify this effect, we show, in the inset, the variance of the conditional distributions of added sizes given birth sizes. The increase in the variance agrees with our extrinsic noise prediction shown in grey: $\sigma^2_\mathrm{fw}[\Delta_d | s_b]=(a s_b + b)^2 \sigma_e^2 $ where $a$ and $b$ were obtained with a linear fit of division sizes against birth sizes. We performed the same analysis with the other datasets, and show the added size distributions for \textit{E. coli} in LB medium at \SI{27}{\degree C} in \cref{fig:data_noise}e, and for \textit{M. smegmatis} in glycerol in \cref{fig:data_noise}f (see SM for other conditions). Similarly, there is no collapse of the conditional added-size distributions, and the increase in variance is consistent with the extrinsic noise model.

We emphasize the importance of accounting for extrinsic noise in theoretical models. Indeed, making the simplifying assumption that noise is additive when it is actually extrinsic in the data leads to a significant relative error between the standard deviations of added size for small-born and big-born cells, ranging from $21\%$ to $28\%$ for analysed \textit{E. coli} data (see SM). In comparison, modelling the mechanism of cell size control as an adder when the experimental slope $a$ is close to but different from $1$ leads to a relative error between the mean added sizes for small-born and big-born cells ranging only from $-4\%$ to $12\%$ for the same data.

\subsubsection{Predictions of cell size control in populations agree with fitted models}

\begin{figure}
    \centering
    \includegraphics[width=\linewidth]{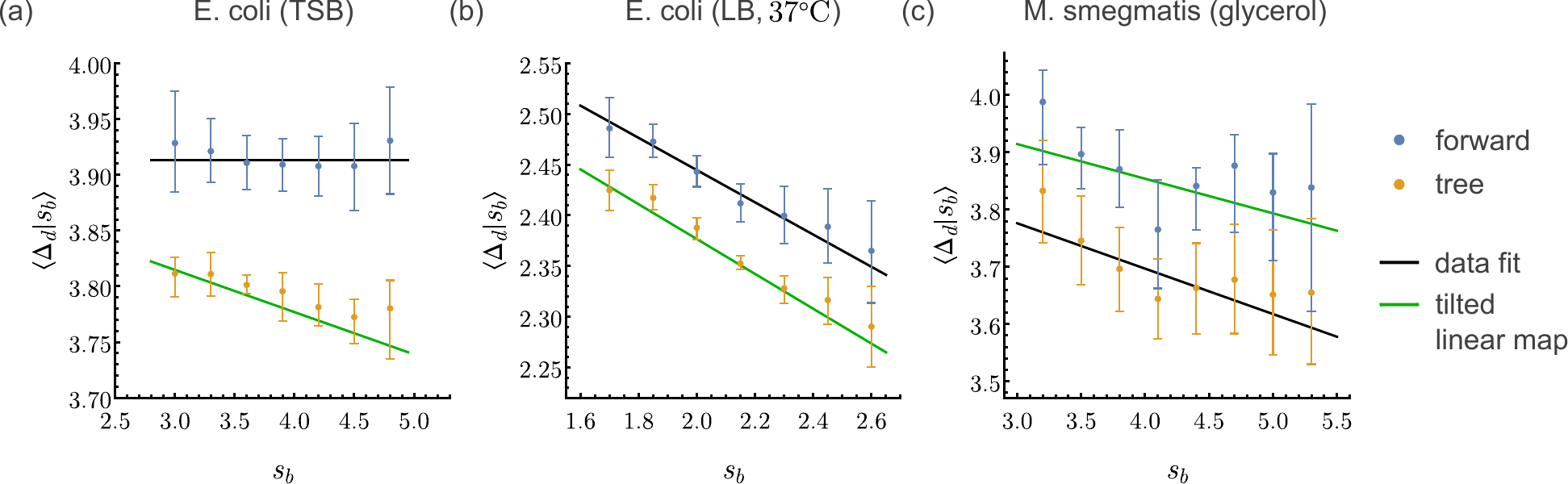}
    \caption{\textbf{Predictions for cell size control using data from a different experimental setup.} (a,b) Mean added size against birth size for \textit{E. coli} in mother machines (blue) vs prediction of \cref{powell_cond_sd} for the same quantity in growing population experiments (orange). Error bars denote the $95\%$ confidence interval of the estimate determined by bootstrapping. The linear fits of the original data are shown in black, and the predictions of the tilted linear map, \cref{eq_pop_lin_map}, with fitted parameters are shown in green. (c) For \textit{M. smegmatis}, population-level data are used to compute forward predictions using \cref{powell_cond_sd}, and using \cref{eq_pop_lin_map} with fitted parameters. Because population datasets are much smaller, error bars denote the $80\%$ confidence interval of the bootstrap. To reduce noise, averages were computed with sliding size windows of width (a) 1, (b) 0.3, and (c) 1.}
    \label{fig:data_slope_pred}
\end{figure}

We now show how cell size control would be modulated if \textit{E. coli} cells observed in mother machines were grown in populations instead. 
In \cref{fig:data_slope_pred}a,b, we show in blue the mean conditional added size against the birth size in the forward statistics for \textit{E. coli} in TSB and in LB at \SI{37}{\degree C}. 
We predict the mean conditional added size in the tree statistics in both conditions using \cref{powell_cond_sd} and kernel density estimations of the conditional forward distribution for different birth sizes. The predictions, shown in orange, are therefore non-parametric and do not assume any mechanism of cell size control.
The black lines show the best linear fits of the unbinned forward data, and the red lines indicate theoretical predictions of the population-level linear map $\langle \Delta_d | s_b \rangle_\mathrm{tree}=(\hat{a}-1)s_b+\hat{b}$ computed with \cref{eq_pop_lin_map} and with parameters from the fitted noise models. 
This shows that the tilted linear map predictions with fitted parameters agree very well with the non-parametric predictions of \cref{powell_cond_sd}.
Importantly, \cref{powell_cond_sd,eq_pop_lin_map} can be inverted to predict forward statistics given statistics obtained in growing population experiments. We show these predictions in \cref{fig:data_slope_pred}c for \textit{M. smegmatis} in glycerol. 
We observe that in TSB the population slope $\hat{a}$ is clearly modulated, producing a stronger mechanism of cell size control (closer to the sizer) at the population level, while in the two other conditions the modulation enters mainly through the intercepts, such that target added sizes are smaller at the population level.

\subsection{Growth rate modulation of cell size control}
\label{sec_grow_rate}

\begin{figure*}
    \includegraphics[width=\linewidth]{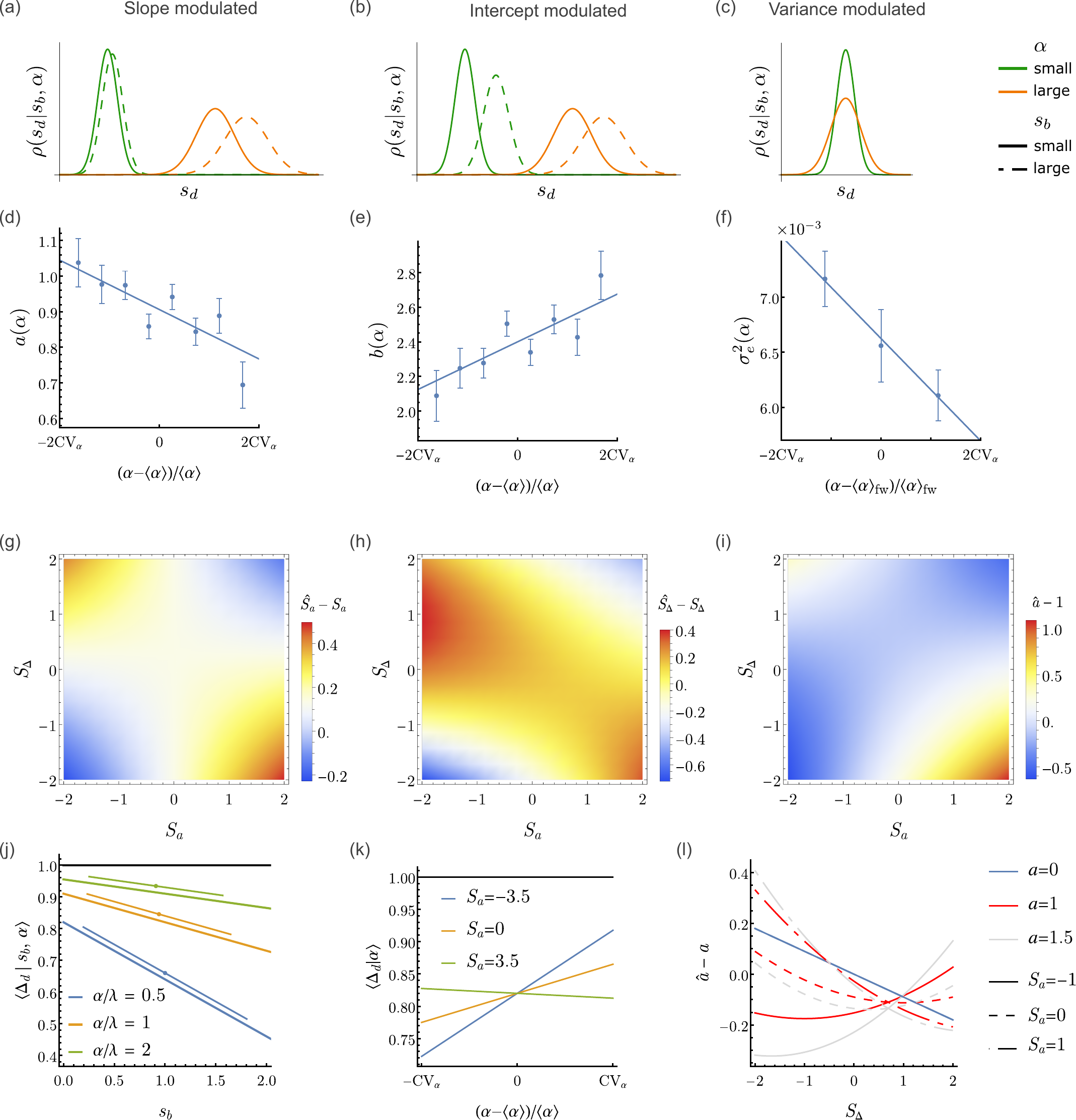}
    \caption{\textbf{Growth rate modulation affects cell size control both in lineages and populations.} (a-c) Modes of growth rate modulation of cell size control occur through slope, intercept, and variance. (d-f) Data suggest all cell size control variables are modulated for \textit{E. coli} in glucose. 
    Slopes $a(\alpha)$ and intercepts $b(\alpha)$ were obtained with linear fits of $s_d$ against $s_b$ for single-cell growth rates $\alpha$ in the interquantile interval $[2.5\%,97.5\%]$ divided into eight bins, and extrinsic variances $\sigma_e^2(\alpha)$ by fitting $\sigma^2[s_d|s_b,\alpha]$ against $\langle s_d|s_b,\alpha \rangle$ when varying $s_b$, for $\alpha$ in the same interval divided into three bins.
    Error bars represent the standard errors on the estimated parameters.
    (g-i) The model predicts how the lineage-population biases in sensitivities and in slope depend on cell size control variables and added size sensitivity. $\mathrm{CV}_\alpha=0.4$, $\sigma_e=0.3$. (g) $S_\sigma=-0.5$. (h) $a=1$, $\mathrm{CV}_p=0$. (i) $a=1$. (j,k) Even without growth rate modulation of a forward adder (black, $a=1$), population size control variables depend on single-cell growth rate. (j) $S_a=S_b=S_\sigma=0$, $\sigma_e=0.3$. (k) $\mathrm{CV}_p=0$, $S_\Delta=0$, $\sigma_e=0.3$. (l) Lineage-population bias in size control mechanisms differs between the adder (red), sizer (blue) and timer-like (grey). $\mathrm{CV}_\alpha=0.4$, $\sigma_e=0.3$.}
    \label{fig:growth_rates}
\end{figure*}

Cell size control can, in principle, be modulated by any cell property, like single-cell growth rates. While the dependence of added size on growth rate for the adder mechanism has been investigated \cite{modi_analysis_2017,grilli_empirical_2018}, a more thorough analysis of the modulation of the noisy linear map by growth rates has been lacking. Recently, the authors have shown that slope $a(\alpha)$ and intercept $b(\alpha)$ depend on growth rate for \textit{E. coli} in mother machines \cite{genthon_noisy_2025}. Here, we further explore the impact of fluctuating growth rates on the noisy part of map. Growth rate modulations of the slope, intercept and variance are qualitatively different, as illustrated in \cref{fig:growth_rates}a-c. 

Analysis of \textit{E. coli} data revealed that noise is extrinsic for most conditions and most single-cell growth rates (see SM), and that the variance of the extrinsic noise depends on single-cell growth rates. We therefore focus on the analysis of the extrinsic noise model, with leading order dependencies of the slope, intercept and variance on single-cell growth rate
around the mean given by:
\begin{subequations}
    \begin{align}
    a(\alpha) &= a \lbk 1  + S_a \frac{\alpha - \langle \alpha \rangle_\mathrm{fw}}{\langle \alpha \rangle_\mathrm{fw}} \rbk \\
    b(\alpha) &= b \lbk 1  + S_b \frac{\alpha - \langle \alpha \rangle_\mathrm{fw}}{\langle \alpha \rangle_\mathrm{fw}} \rbk \\
    \sigma^2_e(\alpha) &= \sigma_e^2 \lbk 1  + S_{\sigma} \frac{\alpha - \langle \alpha \rangle_\mathrm{fw}}{\langle \alpha \rangle_\mathrm{fw}} \rbk \,,
\end{align}
\end{subequations}
where $a=\int a(\alpha) \rho_\mathrm{fw} (\alpha) d\alpha$, $b=\int b(\alpha) \rho_\mathrm{fw} (\alpha) d\alpha$ and $\sigma_e^2=\int \sigma^2_e(\alpha) \rho_\mathrm{fw} (\alpha) d\alpha$ are the slope, intercept and extrinsic variance averaged over growth rates; and $S_a$, $S_b$ and $S_\sigma$ are the sensitivities of slope, intercept and extrinsic variance on growth rate. 
We show these dependencies in \cref{fig:growth_rates}d-f for \textit{E. coli} in glucose (see SM for other growth media and temperatures).

For simplicity, we assume no mother-daughter correlations in single-cell growth rates, yet mother-daughter correlations in sizes and generation times are present as a result of size control \cite{lin_effects_2017,genthon_noisy_2025}. 
In this case, growth rates and birth sizes are independent in the forward statistics, and integrating the linear map over birth sizes, using $\langle s_b \rangle_\mathrm{fw} = \langle \Delta_d \rangle_\mathrm{fw} = b/(2-a)$, reveals that added size depends on growth rate as $\langle \Delta_d | \alpha \rangle_\mathrm{fw} = \langle \Delta_d  \rangle_\mathrm{fw} \lbk 1 + S_\Delta (\alpha - \langle \alpha \rangle_\mathrm{fw})/\langle \alpha \rangle_\mathrm{fw} \rbk$ with added size sensitivity
\begin{equation}
\label{eq_Del_sensi}
    S_\Delta = a S_a + (2-a) S_b \,.
\end{equation}
Values of $S_\Delta$ obtained with direct fits of added sizes vs growth rates are in excellent agreement with those computed with \cref{eq_Del_sensi} (see SM), thus validating the independence assumption.

\subsubsection{Lineage-population bias}

We now investigate the role of fluctuating growth rates in the lineage-population bias for cell size control studied in \cref{sec_lin_pop_bias}. 
In the small noise regime, this bias [\cref{powell_cond_sd}] reveals a titled noisy linear map at the population level: $s_d = (\hat{a}(\alpha) s_b + \hat{b}(\alpha))(1+\hat{\eta}_e(\alpha))$, where forward extrinsic noise remains extrinsic at the population level as observed in the first part of the article. The modified slope and slope sensitivity are given by:
\begin{subequations}
\label{eq_pop_a_Sa}
\begin{align}
\label{eq_a_alpha_pop}
\hat{a}(\alpha)&\approx a \lbk 1-\sigma_e^2 +\lp \ln 2 - \frac{S_\Delta}{2}\rp S_a \mathrm{CV}_{\alpha}^2 + \hat{S}_a \frac{\alpha - \langle \alpha \rangle_{\mathrm{tree}}}{\langle \alpha \rangle_{\mathrm{tree}}} \rbk \\
\hat{S}_a&\approx  S_a  +(1-S_{\sigma}-S_a) \sigma_e^2 + S_a \lp \ln 2 - \frac{S_\Delta}{2}\rp \mathrm{CV}_{\alpha}^2 \,.
\end{align} 
\end{subequations}
The modified intercept $\hat{b}(\alpha)$ and intercept sensitivity $\hat{S}_b$ follow the same transformations as for the slope and are given by \cref{eq_pop_a_Sa} when replacing $a$ by $b$ and $S_a$ by $S_b$; and the modified noise is given in the SM.

In addition to the modulation through extrinsic noise already analysed in \cref{sec_lin_pop_bias}, the mechanism of cell size control is also modulated through noise in single-cell growth rates. The latter either increases or decreases the mode $a$ of size control depending on the sign of $\lp \ln 2 - S_\Delta/2\rp S_a$. 
The slope sensitivity on growth rate is also modulated by extrinsic noise and growth rate noise. 
We show in \cref{fig:growth_rates}g that slope sensitivity can be either reduced or increased in the population statistics. 
Importantly, even when size control and single-cell growth are uncoupled in the forward dynamics ($S_a=S_b=S_\sigma=0$), the size control mechanism becomes sensitive to single-cell growth at the population level with $\hat{S}_a=\sigma_e^2$. 
This is a consequence of the interplay between selection occurring in growing populations and fluctuations in single-cell growth rates, which generates correlations between birth sizes and single-cell growth rates.
We illustrate this point in \cref{fig:growth_rates}j with a growth-rate-independent forward adder, where $b+\eta_e$ follows a Gamma-distribution of mean $b=1$ and standard deviation $\sigma_e=0.3$. Population-level size control is computed using \cref{powell_cond_sd} for different scaled single-cell growth rates $\alpha/\lambda$ shown in different colours. The predictions of the tilted linear map \cref{eq_pop_a_Sa} are shown on an interval of two standard deviations below and above the mean birth size, which is also modulated by growth rates and given in the SM, indicated by a circle. We observe that
slow-growing cells deviate towards the sizer more than fast-growing cells. 

As a consequence of the tilted linear map, the average added size is also modulated in a growth-rate-dependent manner in the tree statistics. For a forward adder, the mean added size and its sensitivity on growth rate are given by (see SM for details and for expression with general $a$):
\begin{subequations}
    \begin{align}
    \langle \Delta_d | \alpha \rangle_{\mathrm{tree}} &\approx \langle \Delta_d  \rangle_{\mathrm{fw}} \lbk 1 - 2 \sigma_e^2 + \lp \ln 2 -\frac{S_\Delta}{2} \rp S_\Delta \mathrm{CV}_{\alpha}^2 +\hat{S}_\Delta \frac{\alpha-\langle\alpha\rangle_{\mathrm{tree}}}{\langle\alpha\rangle_{\mathrm{tree}}} \rbk \\ 
    \label{eq_S_Delta_pop}
    \hat{S}_\Delta &\approx  S_\Delta + \frac{4 S_a}{3} \mathrm{CV}_p^2 + \frac{6-2 S_a - 3 S_\Delta}{3} \sigma_e^2 + \lbk S_a \lp \frac{S_\Delta}{6}-\ln 2 \rp+ \ln 2 -\frac{S_\Delta}{2} \rbk S_\Delta \mathrm{CV}_{\alpha}^2 \,.
\end{align}
\end{subequations}
The added size sensitivity can be both increased or decreased at the population level, as illustrated in \cref{fig:growth_rates}h.
Again, even when the forward mechanism of cell size control is independent of single-cell growth ($S_a=S_b=0$), added size becomes an increasing function of single-cell growth rate at the population level with sensitivity $\hat{S}_\Delta=2\sigma_e^2$. 
This result is in qualitative agreement with the increase of added size with single-cell growth rate in population statistics, reported for \textit{E. coli} in M9 medium \cite{campos_constant_2014} and for \textit{M. smegmatis} \cite{priestman_mycobacteria_2017},
although the sensitivity induced by extrinsic noise alone is too small to quantitatively explain these behaviours.
More generally, added size can also become a decreasing function of single-cell growth rate at the population level, even when they are independent in forward lineages ($S_\Delta=0$). This regime where $\hat{S}_\Delta \leq 0$ can be reached either with large positive or large negative values of $S_a$, depending on the sign of $2\mathrm{CV}_p^2 - \sigma_e^2$. The different trends of added size vs single-cell growth rate when varying $S_a$ for $\mathrm{CV}_p=0$ are shown in \cref{fig:growth_rates}k.

In most studies on cell size control, single-cell growth rates are not measured and the mechanism of size control is inferred by fitting division sizes against birth sizes. 
We therefore integrate out growth rates to derive the population-level titled linear map relating division size to birth size irrespective of growth rate, with modified slope and intercept (see SM for derivation and modified noise term):
\begin{subequations}
    \begin{align}
    \label{eq_a_pop_alpha_int}
    \hat{a}&\approx a\lbk 1- \sigma_e^2 + \lp \frac{S_\Delta(S_\Delta+2)}{4} + S_a (\ln 2 - S_\Delta)\rp \mathrm{CV}_{\alpha}^2  \rbk - S_\Delta \mathrm{CV}_{\alpha}^2 \\
    \hat{b}&\approx b\lbk 1- \sigma_e^2 + \lp \frac{S_\Delta(S_\Delta+2)}{4} + S_b (\ln 2 - S_\Delta)\rp \mathrm{CV}_{\alpha}^2  \rbk \,.
\end{align}
\end{subequations}
While extrinsic noise reduces the slope and the intercept at the population level, growth rate noise can either increase of decrease them. We show the modulation of the mode of cell size control for a forward adder in \cref{fig:growth_rates}i. 
Unlike in previous results \cref{eq_pop_lin_a,eq_a_alpha_pop}, a sizer in forward lineages is not preserved at the population level, which is a consequence of the coarse-graining over single-cell growth rates (see SM). We show how the sizer, the adder and a timer-like mechanism are differently modulated through growth rate noise in \cref{fig:growth_rates}l.

\subsubsection{Trade-off between growth rate gain and noise minimisation}

\begin{figure}
    \centering
    \includegraphics[width=\linewidth]{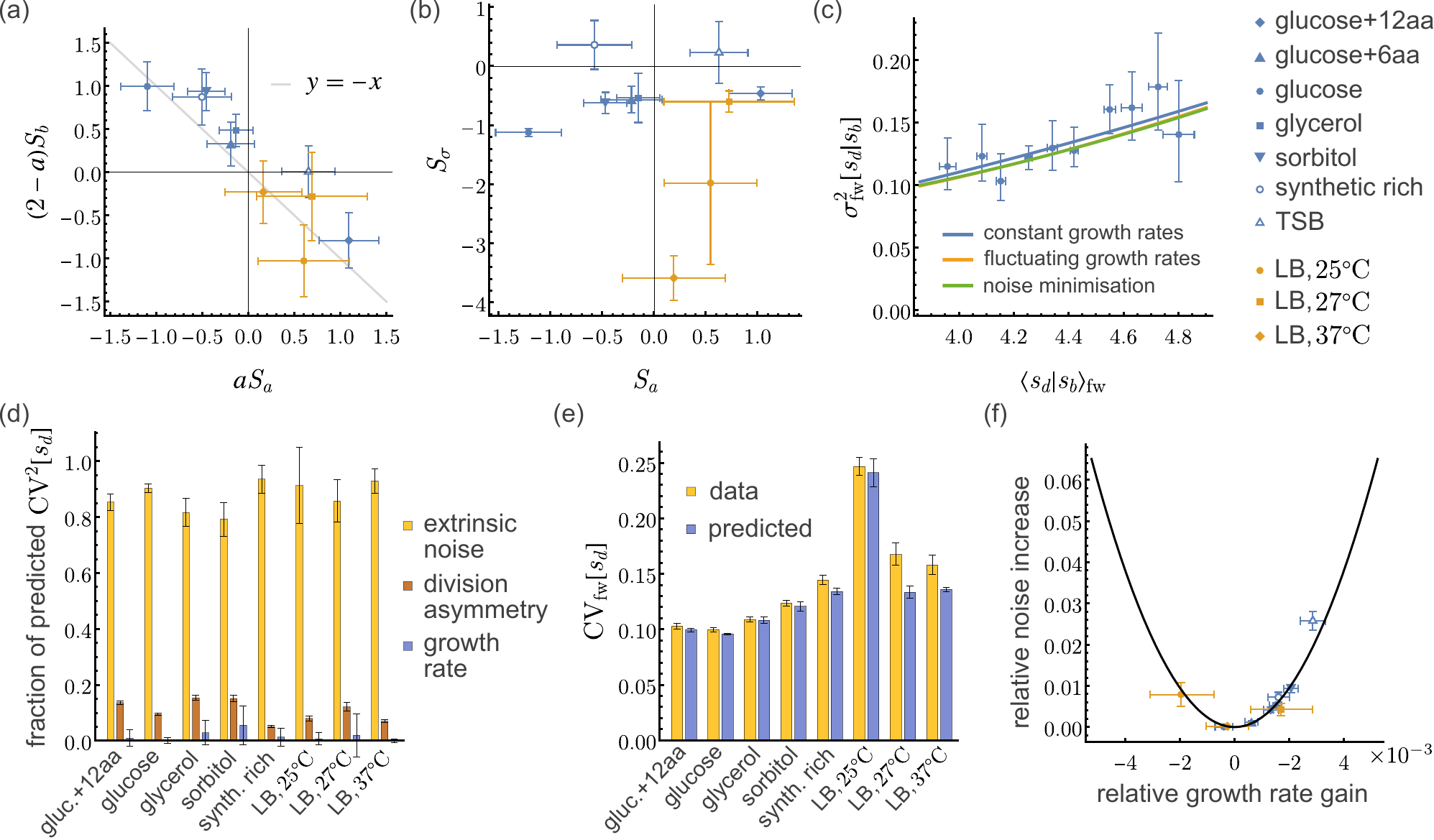}
    \caption{\textbf{\textit{E. coli} optimises noise in division size instead of growth rate gain.} (a) Correlation between slope and intercept sensitivities along the line $S_\Delta=0$ (grey) across growth media and temperature conditions. (b) No correlation between slope and variance sensitivity. (a,b) Error bars indicate the standard errors on the estimates provided by the best fits of slope and intercept against growth rates.
    (c) Different noise models for \textit{E. coli} in glucose: extrinsic noise when ignoring growth rate fluctuations (blue), model with growth-rate noise \cref{var_alpha_v2} (orange), and minimised growth-rate noise ($S_\Delta=0$) (green). All models agree with data, i.e. mean-variance relation of division size is not influenced by growth rate variability due to noise minimisation. See \cref{fig:data_noise} for error bars.
    (d) Contributions of extrinsic noise are the dominant source of total noise in cell size across all growth conditions.
    (e) Total noise in division sizes from data is consistent with our model prediction across growth conditions.
    (f) Data lie close to the minimum of the trade-off curve between the relative noise increase, defined as the fraction of the total CV in division size caused by fluctuating growth rates, and the relative growth rate gain, defined as the growth rate gain due to size control sensitivities relative to the population growth rate when $S_\Delta=0$, computed according to \cref{eq_CV_sd} and \cref{eq_lambda} respectively. 
    }
    \label{fig:noise_min}
\end{figure}

We have seen that cell size control is modulated by single cell growth at the lineage and population levels, however what sets the values of the size control sensitivities on growth rates remains unknown. 
Here, we show that sensitivity values impact both the accuracy of size control and population growth, which reveals a trade-off.

Fluctuating growth rates can be interpreted as an extra source of noise on division size. Indeed, the conditional variance of division size when growth rate is integrated out is given by the law of total variance at leading order: 
\begin{equation}
\label{var_alpha_v2}
    \sigma_{\mathrm{fw}}^2[s_d|s_b]\approx (a s_b+b)^2 \sigma_e^2 +\lbk a S_a(s_b-\langle s_b \rangle_{\mathrm{fw}}) + S_\Delta \langle s_b \rangle_{\mathrm{fw}}  \rbk^2 \mathrm{CV}_{\alpha}^2 \,,
\end{equation}
where the second term corresponds to the variance \textit{explained} by the differences in mean division sizes for different growth rates, $\sigma^2[\langle s_d | s_b,\alpha\rangle|s_b]$, and where the first term is the \textit{unexplained} variance $\langle \sigma^2[s_d|s_b,\alpha]| s_b\rangle$ (neglecting next order terms in $\sigma_e^2 \mathrm{CV}_{\alpha}^2$) coming from division size variability between cells with the same growth rates.
Fluctuating growth rates thus generate a combination of additive, intrinsic and extrinsic noises (identifiable when expanding the bracket in \cref{var_alpha_v2} and grouping terms in powers of $a s_b+b$). This combination simplifies in some limits: when $S_a=0$ it reduces to an additive noise with variance $\sigma_a^2=S_\Delta^2 \langle s_b \rangle_{\mathrm{fw}}^2  \mathrm{CV}_{\alpha}^2$; and when $S_a=S_b \equiv S$ it reduces to an extra extrinsic noise with variance $\Tilde{\sigma}_e^2=S^2 \mathrm{CV}_{\alpha}^2$. 

The growth-rate-induced noise on conditional division size can be minimised when the bracket in \cref{var_alpha_v2} is small, that is when $S_\Delta=0$, leaving a term which is small close to the mean birth size. Similarly, we show that the marginal variability in division sizes is given by (see SM):
\begin{equation}
\label{eq_CV_sd}
    \mathrm{CV}_{\mathrm{fw}}^2[s_d]\approx\frac{a^2}{4-a^2}\mathrm{CV}_{p}^2 + \frac{4}{4-a^2}\sigma_e^2 + \frac{S_\Delta^2}{4-a^2} \mathrm{CV}_{\alpha}^2 \,.
\end{equation}
As with the conditional division size variability, the contribution from fluctuating growth rates is minimised when $S_\Delta=0$.

On the other hand, it is well understood that population growth depends on single-cell growth rate statistics. Uncorrelated fluctuations in single-cell growth rates are detrimental for population growth rate $\lambda$ when single-cell growth and size control are uncoupled \cite{thomas_single-cell_2017,lin_effects_2017}, but can become beneficial when size control is sensitive to single-cell growth \cite{genthon_noisy_2025}:
\begin{equation}
\label{eq_lambda}
    \frac{\lambda}{\langle \alpha \rangle_\mathrm{fw}} \approx 1 - \lbk 1 - \frac{\ln2}{2} - \frac{S_\Delta}{2 \ln 2} \rbk \mathrm{CV}_{\alpha}^2 \,.
\end{equation}
The population growth rate increases with $S_\Delta$, implying a trade-off between population growth and size-control accuracy. 

To test where \textit{E. coli} stands in this trade-off, we report in \cref{fig:noise_min}a the values of $a S_a$ and $(2-a) S_b$ obtained in different growth media and temperatures. We observe that for all conditions, these two quantities are of opposite signs, and that they align with the $S_\Delta=0$ line shown in grey. 
This could indicate that size control sensitivities $S_a$ and $S_b$ on single-cell growth are regulated to minimise size fluctuations in \textit{E. coli} instead of maximising population growth. 
Note that the extrinsic variance sensitivity $S_\sigma$ plays no role at leading order in this trade-off. This is consistent with the fact that values of $S_\sigma$ plotted against $S_a$ in \cref{fig:noise_min}b do not show any clear trend, indicating that $S_\sigma$ might not be tightly regulated.

To quantitatively test noise minimisation, we show the contribution of growth rate fluctuations to both conditional and marginal division size variability in \cref{fig:noise_min}c,d. 
In \cref{fig:noise_min}c, we compare the extrinsic fit obtained in the first part of the paper when ignoring growth rate fluctuations, in blue; the prediction of \cref{var_alpha_v2} with fitted size control parameters, in orange; and the prediction obtained when minimising growth-rate-induced noise by setting $S_\Delta=0$ in \cref{var_alpha_v2}, in green; for \textit{E. coli} in glucose (see SM for other conditions). All three noise models are consistent with the data within the error bars. The proximity between the blue and orange curves indicates that extrinsic noise dominates growth-rate-induced noise, and the superimposition of the green and orange curves shows that \textit{E. coli} is close to the regime of noise minimisation. 
In \cref{fig:noise_min}d, we show the decomposition of the division size variability into contributions coming from extrinsic noise on size control, division asymmetry and single-cell growth rates using \cref{eq_CV_sd}. For all conditions, extrinsic noise again dominates, noise in division asymmetry contributes at most $15\%$, and growth rate noise is negligible, consistent with the regime of noise minimisation.
To test the validity of our model prediction, \cref{eq_CV_sd}, we compared it to the coefficients of variation measured in \textit{E. coli} data in \cref{fig:noise_min}e, and observed a very good agreement for all conditions except in LB medium at \SI{27}{\degree C} and \SI{37}{\degree C} where the model slightly underestimates variability. 

Finally, we show where \textit{E. coli} data stand on the trade-off curve between noise minimisation and growth rate gain in \cref{fig:noise_min}f. The y-axis represents the fraction of the total variability in division size caused by fluctuating growth rates, $S_\Delta^2 \langle \mathrm{CV}_{\alpha}^2 / ((4-a^2)\mathrm{CV}_{\mathrm{fw}}^2[s_d]) \rangle_\mathrm{cond}$, and the x-axis represents the growth rate gain due to size control sensitivities relative to the population growth rate when sensitivities are null, $S_\Delta \langle \mathrm{CV}_{\alpha}^2 / ((2 \ln 2)(1-(1-\ln 2 /2)\mathrm{CV}_{\alpha}^2)) \rangle_\mathrm{cond}$, where $\langle \rangle_\mathrm{cond}$ indicates an average over growth conditions. We used the estimate of $S_\Delta$ obtained from the direct fit of added sizes versus growth rates for the noise increase, and the value of $S_\Delta$ computed using \cref{eq_Del_sensi} for the growth-rate gain. This shows that all data are close to the minimum of the trade-off curve, with negligible gains in growth rate to regulate noise in division size.

\section{Discussion}

Cellular variation in birth and division size allows us to study how cell size control occurs in living cells. In this work, we show that extrinsic noise, growth rate, and population dynamics jointly shape the apparent distribution of size control measured in experiments. We present two stochastic maps that unify phenomenological models of cell size control: one applicable to mother machine lineages and incorporating growth-rate modulation and noise mechanisms, and another, a tilted linear map, applicable to the corresponding lineage trees of cell populations. We demonstrate that widely used quantifiers of the division-size distribution, such as the slope of the birth-division-size relation and the added size, depend on the experimental context and growth rate. 
Our results identify extrinsic noise as the dominant source of size variability across organisms and conditions, and reveal a trade-off between growth-rate gain and division-size noise that appears to constrain bacterial physiology.

Our analysis reveals modulation of cell-size control by selection effects in exponentially growing populations. This modulation reveals, to leading order, that population-level size control follows a tilted linear map with parameters biased relative to those of the linear map measured in mother machines. Importantly, in growing populations, the adder behaviour becomes more sizer-like in the presence of extrinsic noise, but more timer-like in the presence of additive noise. This contrasts sharply with studies where size-additive and time-additive noises have qualitatively equivalent impacts \cite{jafarpour_cell_2019,lin_effects_2017,kar_are_2025}. This selection-induced bias in size control must be accounted for in data analysis, as it can obscure the underlying size control mechanism, in the same way as survivor biases in experiments with cell death \cite{genthon_cell_2023}, as inappropriate geometric measures of cell size \cite{facchetti_reassessment_2019,kar_are_2025}, and circadian clock modulation of cell size control \cite{martins_cell_2018}.

A key finding of our study is that an extrinsic noise model quantitatively captures division-size fluctuations across a range of data. For an adder, the model predicts that, on average, the added size is independent of birth size, whereas fluctuations generally increase with birth size, as we confirmed in both \emph{E. coli} and \emph{M. smegmatis} in various growth conditions. Conditional variance-mean relationships collapse onto a universal quadratic curve for most growth conditions, indicating multiplicative fluctuations in division size instead of the commonly assumed additive noise. The few conditions that deviate from this collapse, associated with very rich or slow-growth environments, suggest either correlated noise sources or additive noise, highlighting physiological regimes where additional regulation may operate.

We revealed that, in \emph{E. coli}, the mechanism of cell size control depends on cell-growth rate. This means that cells growing either faster or slower than average can deviate significantly in their control of cell size. For example, adder cells can behave more sizer-like or timer-like depending on their growth rate. Moreover, we show that such growth-rate-dependent cell-size control can be observed in population statistics even when mother machine lineages are not growth-rate-dependent. 
Fluctuations in single-cell growth rates thus generate an additional source of division-size noise that alters the apparent cell size control through selection at the population level. 

Importantly, the classical time-additive noise model and the extrinsic framework are equivalent in the weak noise limit for exponentially-growing cells with constant growth rate, since they both produce size-multiplicative noise with the same scaling.
However, time-additive noise, in its original form and variations thereof \cite{amir_cell_2014,hein_asymptotic_2024}, imposes fixed sensitivities of cell size control coefficients ($S_a=S_b=0$, $S_\sigma=2$ in \cite{amir_cell_2014} and $S_\sigma=0$ in \cite{hein_asymptotic_2024}, see SM) that are incompatible with measurements in \emph{E. coli} across conditions (\cref{fig:noise_min}a,b). The general extrinsic noise mechanism with condition-dependent sensitivities to growth rate therefore better captures the size control underlying the data (\cref{fig:noise_min}).

Our results appear to contradict the threshold accumulation models of size control. In such models, division is triggered when a hypothetical division protein reaches a fixed threshold, naturally producing adder behaviour and added size distribution independent of birth size \cite{ghusinga_mechanistic_2016,pandey_exponential_2020} and replication status \cite{si_mechanistic_2019}. This interpretation is more consistent with additive noise models and thus cannot account for the observed extrinsic noise scaling and the absence of collapse in added-size distributions. While we found that extrinsic noise accounts for most datasets, slow-growth conditions are incompatible with this model (see SM for analysis of \textit{E. coli} data in slow-growth condition M9+glycerol from \cite{nieto_mechanisms_2024}) and more closely resemble additive noise predicted by protein threshold accumulation. However, deviations from adder behaviour become most apparent during slow growth, suggesting that additional regulatory layers become relevant under these conditions \cite{nieto_mechanisms_2024}. How threshold-accumulation models and extrinsic noise can be reconciled within a unified mechanistic framework remains an open question \cite{elgamel_effects_2024,luo_stochastic_2023}.

The small-noise expansion allows us to directly compare these mother machine and population setups, but it is strictly valid only in the weak-noise regime, for approximately linear size-control maps and constant growth rates within a cell cycle, without intergenerational correlations. These assumptions enable analytical tractability of the effective cell size statistics and lineage-population bias but may fail in regimes with strong nonlinear regulation \cite{zhang_nonlinear_2025} or considerable physiological heterogeneity. In these cases, the full tilted distributions must be used (\cref{powell_cond_sd}), which generally predict nonlinearity even when the corresponding mother machine map is linear (fig.~\ref{fig:pop_alpha_cst}). Nevertheless, we observed good agreement with the tilted linear map (\cref{fig:data_slope_pred}), validating our assumptions for the data analysed here. 
Finally, our model does not include unobserved molecular variables and thus cannot identify the biochemical origins of noise or compensation. Experiments combining single-lineage and population measurements \cite{ziegler_scaling_2024,bakshi_tracking_2021} with direct molecular readouts of division proteins, such as FtsZ \cite{si_mechanistic_2019}, will be required to extend this framework and resolve the mechanisms underlying these deviations.

Finally, we uncover cell size control in \emph{E. coli} implements a trade-off between population growth-rate gain and size-noise minimisation. While growth-rate-dependent size control can enhance population growth, it inevitably amplifies size variability. Strikingly, across all examined conditions, slope and intercept sensitivities align close to the regime that minimises growth-rate-induced noise, placing cells near the minimum of the predicted trade-off curve. This indicates that bacterial size control prioritizes the robustness of division size and a narrow cell-size distribution over maximising the gain in growth rate. 

Our results reveal that cell size control emerges from the coupling of noise, growth-rate variability, and population dynamics. Dominant extrinsic noise reveals a trade-off where bacteria operate near an optimal regime that prioritizes tight size control over maximal growth. How this trade-off translates to fitness in changing environments remains an open question \cite{martins_cell_2018,panlilio_threshold_2021,nguyen_distinct_2021,bakshi_tracking_2021}.

\section{Methods}

We highlight here the steps to derive the population-level tilted linear map \cref{eq_pop_lin_map} from the tilted distribution of division sizes derived in the SM:
\begin{equation}
    \label{powell_cond_sd}
\rho_{\mathrm{tree}}(s_d|s_b,\alpha) = \frac{2  \rho_{\mathrm{fw}}(s_d|s_b,\alpha) e^{-\lambda \tau_d(s_d,s_b,\alpha)}}{Z(s_b,\alpha)} \,,
\end{equation}
where $Z(s_b,\alpha)$ is a normalization factor, $\lambda$ is the steady-state population growth rate, and $\tau_d(s_d,s_b,\alpha)$ is the interdivision time for a cell with birth size $s_b$, division size $s_d$ and growth rate $\alpha$.
For exponentially-growing cells, $\tau_d(s_d,s_b,\alpha)=\ln(s_d/s_b)/\alpha$, with constant growth rates which imposes $\lambda=\alpha$ \cite{lin_effects_2017}, \cref{powell_cond_sd} takes the form 
\begin{equation}
\label{eq_powell_methods}
\rho_{\mathrm{tree}}(s_d|s_b) = \frac{s_d^{-1}  \rho_{\mathrm{fw}}(s_d|s_b) }{\langle s_d^{-1} | s_b \rangle_\mathrm{fw}} \,,
\end{equation}
where we used the normalization of probability distribution $\rho_{\mathrm{tree}}$ to expressed the factor $Z(s_b)/(2 s_b)=\langle s_d^{-1} | s_b \rangle_\mathrm{tree}$. The conditional mean division size is therefore given by $\langle s_d |s_b \rangle_\mathrm{tree} = 1/\langle s_d^{-1} | s_b \rangle_\mathrm{fw}$. 
Using the forward map $s_d = f(s_b) + \eta$, an expansion of the division size around its mean $f(s_b)$ gives $s_d^{-1} = f(s_b)^{-1} \sum_{n=0}^{\infty} (-\eta/f(s_b))^n$. Because $\eta$ is a zero-mean noise, we then have $\langle s_d^{-1} | s_b \rangle_\mathrm{fw} = f(s_b)^{-1} \lbk 1 + \sum_{n=2}^{\infty} (-1)^n \langle \eta^n | s_b \rangle_\mathrm{fw}/f(s_b)^n \rbk$. 

First, motivated by the small variability in division sizes, we keep only the leading order term in this expansion: $\langle s_d^{-1} | s_b \rangle_\mathrm{fw} = f(s_b)^{-1} \lbk 1 + (\sigma/f(s_b))^2 +o((\sigma/f(s_b))^2)\rbk$, where $\sigma^2=\langle \eta^2 | s_b \rangle_\mathrm{fw}$. The population average is therefore given by $\langle s_d |s_b \rangle_\mathrm{tree}=f(s_b) \lbk 1 - (\sigma/f(s_b))^2 +o((\sigma/f(s_b))^2)\rbk$. Using the independence of the additive, intrinsic and extrinsic noises, we have for the variance $\sigma^2=\sigma_a^2 + f(s_b) \sigma_i^2 + f(s_b)^2 \sigma_e^2$. For a linear map $f(s_b)=a s_b +b$, the population average reads:
\begin{equation}
\label{eq_powell_exp}
    \langle s_d | s_b \rangle_{\mathrm{tree}} \approx (as_b+b) \lbk 1 -  \frac{\sigma_a^2}{(as_b +b)^2} - \frac{\sigma_i^2}{as_b +b} - \sigma_e^2 
    \rbk\,,
\end{equation}
From \cref{eq_powell_exp}, we understand that intrinsic and extrinsic noises maintain a linear dependence between birth size and target division size at the population level. 
This is only true at leading order for intrinsic noise, as the expansion generates non-linear terms in $s_b$ at order $\sigma_i^3$ (see SM). 
With extrinsic noise on the other hand, it is always true that the functional dependence $f(s_b)$ of target division size on birth size is conserved (and scaled), as observed in \cref{fig:pop_alpha_cst}f, since $\langle s_d|s_b \rangle_{\mathrm{tree}}= f(s_b)/\langle (1 + \eta_e)^{-1} \rangle_{\mathrm{fw}}$ is exact for any $f(s_b)$.
At leading order, intrinsic noise does not modulate the slope and only decreases the intercept by $\sigma_i^2$, while extrinsic noise decreases both the slope and the intercept by $\sigma_e^2$.

Second, to compute the contribution of additive noise to the population-level linear map, we use the small variability in birth sizes, and we linearize \cref{eq_powell_exp} around the mean birth size in the population statistics $\langle s_b \rangle_{\mathrm{tree}}$:
\begin{equation}
\label{eq_powell_exp_2}
    \langle s_d | s_b \rangle_{\mathrm{tree}} \approx a \lbk 1 + \frac{\sigma_a^2}{(a \langle s_b \rangle_{\mathrm{tree}} +b)^2} - \sigma_e^2 \rbk s_b + b \lbk 1 - \frac{(2a \langle s_b \rangle_{\mathrm{tree}} +b)\sigma_a^2}{b(a \langle s_b \rangle_{\mathrm{tree}} +b)^2} - \frac{\sigma_i^2}{b} - \sigma_e^2 \rbk \,.
\end{equation}
We then express $\langle s_b \rangle_{\mathrm{tree}}$, which is derived separately as detailed in the SM and given in \cref{eq_bias_sb}. Since corrections to the mean birth size coming from additive, intrinsic and extrinsic noises on division size would generate higher order corrections in \cref{eq_powell_exp_2}, we can ignore them at this level. We thus express $\langle s_b \rangle_{\mathrm{tree}} = \langle s_b \rangle_{\mathrm{fw}} \lbk 1 + 4 \mathrm{CV}_p^2/(4-a^2) \rbk $ in \cref{eq_powell_exp_2}, to recover the expressions for the population slope and intercept from the main text. 

We understand that the linear map modulation due to noise in division asymmetry arises from the mean birth size in the population statistics, around which the linear map is a valid approximation. Note that terms proportional to $\sigma_i^3 \mathrm{CV}_p^2$ coupling intrinsic noise and noise in division asymmetry appear for both the modified slope and modified intercept in the expansion at the next order. On the other hand, division asymmetry does not modulate the tilted linear map when coupled only to extrinsic noise, a consequence of the linearity of the population-level linear map with extrinsic noise discussed above.

The variance part of the map is similarly obtained by computing $\langle s_d^2 |s_b \rangle_\mathrm{tree} = \langle s_d | s_b \rangle_\mathrm{fw}/\langle s_d^{-1} | s_b \rangle_\mathrm{fw}$ and expanding around the forward mean division size at third order, which involves the skewness of the noises. Again, the observation that forward extrinsic noise remains extrinsic at the population level is a consequence of $\langle s_d^2|s_b \rangle_{\mathrm{tree}} = f(s_b)^2 /\langle (1 + \eta_e)^{-1} \rangle_{\mathrm{fw}} $.

\section*{Funding}
PT was supported by UKRI through a Future Leaders Fellowship (MR/T018429/1 and MR/Y034309/1).


%

\end{document}


\title{Supplementary Material for \\~\\ `Cell size control in bacteria is modulated through extrinsic noise, single-cell- and population-growth'}

\author{Arthur Genthon}
\author{Philipp Thomas}

\maketitle
\tableofcontents
\newpage

\section{Model}

\subsection{Population dynamics}
\label{sec_pop_dyn}

Cells are characterized by their age $\tau$ and their size $s$, which is a catch-all descriptor for the appropriate geometrical dimension: length, surface, or volume. 
Cell size increases at a speed $ \di s/ \di \tau=g(s,\alpha)$, where the single-cell growth rate $\alpha$ can fluctuate from one cell to another but remains constant during the cell cycle. The function $g$ captures different growth modes, like exponential growth when $g(s,\alpha)=\alpha s$, observed for bacterial cells  \cite{wang_robust_2010,robert_division_2014,osella_concerted_2014,campos_constant_2014,taheri-araghi_cell-size_2015}, budding yeast \textit{S. cerevisiae} \cite{soifer_single-cell_2016}, archaeal cell \textit{H. salinarum} \cite{eun_archaeal_2017}, and mycobacteria such as \textit{M. smegmatis} \cite{priestman_mycobacteria_2017}, for example; linear growth when $g(s,\alpha)=\alpha$, observed for 
\textit{M. tuberculosis}
\cite{chung_single-cell_2024}; or even more complex growth laws such as super-exponential growth recently observed for \textit{E. coli} \cite{kar_distinguishing_2021,genthon_analytical_2022,cylke_super-exponential_2023}, asymptotic linear growth of 
\textit{C. glutamicum} 
\cite{messelink_single-cell_2021}, and piecewise growing patterns of fission yeast \textit{S. pombe} \cite{horvath_cell_2013,jia_characterizing_2022}. 

The expected numbers of cells $n(\tau,s_d,s_b,\alpha;t)$ with age $\tau$, division size $s_d$, birth size $s_b$ and single cell growth rate $\alpha$ in the population at time $t$ obeys:
\begin{equation}
	\label{eq:pbe_alpha}
	\partial_t n(\tau,s_d,s_b,\alpha;t) = - \partial_{\tau}  n(\tau,s_d,s_b,\alpha;t) \,,
\end{equation}
for $0 < \tau \leq \tau_d(s_d,s_b,\alpha)$, where $\tau_d(s_d,s_b,\alpha)$ is the generation time (age at division). The generation time is fully characterized by the other cell parameters: $\tau_d(s_d,s_b,\alpha)= \ln(s_d/s_b)/\alpha$ for exponentially-growing cells, and $\tau_d(s_d,s_b,\alpha)= (s_d-s_b)/\alpha$ for linearly-growing cells, for example. Please note that we treat the size at division $s_d$ as a variable for convenience, which is equivalent to formalisms based on division rates \cite{genthon_noisy_2025}.

The boundary condition for newborn cells is given by
\begin{equation}
	\label{eq:bc_alpha}
	n(\tau=0,s_d,s_b,\alpha;t) = 2 \int \di s_d' \di s_b' \di \alpha' \ 
	\mathcal{K}(s_d,s_b,\alpha|s_d',s_b',\alpha') n(\tau_d',s_d',s_b',\alpha';t) \,,
\end{equation}
where 
the division kernel $\mathcal{K}(s_d,s_b,\alpha|s_d',s_b',\alpha')$ is the probability for a newborn cell to inherit a size at birth $s_b$, a size at division $s_d$ and a growth rate $\alpha$ knowing that its mother was born with a size $s_b'$ and a growth rate $\alpha'$ and divided at size $s_d'$. 
In the following, we consider the class of kernels 
\begin{equation}
	\label{eq_K_kernel}
	\mathcal{K}(s_d,s_b,\alpha|s_d',s_b',\alpha') = \psi(s_d|s_b,\alpha)\nu(\alpha)\kappa(s_b|s_d')\,, 
\end{equation}
where $\psi(s_d|s_b,\alpha)$ encodes the mechanism of cell size control for the newborn cell, $\nu(\alpha)$ is the distribution of single-cell growth rates where we assume no mother-daughter correlations, and $\kappa(s_b|s_d')$ regulates the size partitioning between the two daughter cells upon division. 
We further focus on the case where the partitioning of volume at division only depends on the fraction $p=s_b/s_d'$ of the volume of the mother cell inherited by the daughter cell via
\begin{equation}
	\label{eq:def_kappa}
	\kappa(s_b|s_d')= \int_0^1 \di p \ \pi(p) \delta\lp s_b- p s_d' \rp \,.
\end{equation}
The noise in volume partitioning is thus quantified by $\mathrm{CV}_p^2=\lp \langle p^2 \rangle_{\pi} - \langle p\rangle_{\pi}^2 \rp / \langle p\rangle_{\pi}^2$ with $\langle p\rangle_{\pi}=1/2$.

\subsection{Probability distributions in a population tree}

In steady-state, the cell number grows exponentially:
\begin{equation}
	n(\tau,s_d,s_b,\alpha;t) \underset{t \to \infty}{\sim} \Pi(\tau,s_d,s_b,\alpha) e^{\lambda t} \,,
\end{equation}
where $\lambda$ is the population growth rate, or Malthus parameter, and $\Pi(\tau,s_d,s_b,\alpha)$ is the steady-state snapshot joint distribution of age, size at division, size at birth and single-cell growth rate. 

Cells currently in the population, which were born before the observation time but which have not divided yet, are called \textit{leaves}. Different distributions can be defined with population tree data depending on whether or not leaves are included \cite{lin_effects_2017}. 
In this article, we define two distributions at the tree level: the probability density obtained when sampling cells that divided before the observation time (thus excluding the leaves), used in the main text:
\begin{align}
	\rho_{\mathrm{tree}}(s_d,s_b,\alpha) = \frac{ \Pi(\tau_d,s_d,s_b,\alpha)}{\int \di s_d' \di s_b' \di \alpha'  \ \Pi(\tau_d',s_d',s_b',\alpha')} \,,
\end{align}
where we recall that the generation time $\tau_d(s_d,s_b,\alpha)$ is a deterministic function of the other parameters; 
and the probability density obtained when sampling cells that were born before the observation time (thus including the leaves)
\begin{align}
	\rho_{\mathrm{tree}}^{\mathrm{nb}}(s_d,s_b,\alpha) = \frac{ \Pi(0,s_d,s_b,\alpha)}{\int \di s_d' \di s_b' \di \alpha'  \ \Pi(0,s_d',s_b',\alpha')} \,.
\end{align}
The superscript nb for the second distribution stands for newborn and indicates that leaves are included. In the following, we will refer to the first distribution, used in the main text, as the tree distribution; and to the second one as the newborn distribution.

The reason why these two distributions differ is because the number of leaf cells in an exponentially growing population is comparable to the number of non leaf cells. 
Indeed, a colony starting with one ancestor cell and ending up with $N$ leaf cells has $N-1$ non-leaf cells in its population history, meaning that leaf cells contribute to half the newborn distribution for large $N$. Therefore, the newborn distribution is biased compared to the tree distribution towards cells which are more likely to be observed in the snapshot at final time, which are those with larger generation times. 

Note that it is possible to define the marginal distribution $\rho_{\mathrm{tree}}(s_b)$, which is the distribution of birth sizes among dividing cells. Conversely, the marginal distribution $\rho_{\mathrm{tree}}^{\mathrm{nb}}(s_d)$ is the distribution of division sizes among newborn cells, and is by definition inaccessible since those cells have not divided yet.

\subsection{Forward lineages}

In mother machine experiments, colonies are constrained to grow confined in a one-dimensional microfluidic channel, generating single lineages. In this case, the joint probability distribution $p(\tau,s_d,s_b,\alpha;t)$ obtained by following in time the 'mother cell' at the closed end of the channel differs from $\Pi(\tau,s_d,s_b,\alpha)$ and follows the equation
\begin{equation}
	\label{eq_pde_fw}
	\partial_t p(\tau,s_d,s_b,\alpha;t) = - \partial_{\tau}  p(\tau,s_d,s_b,\alpha;t) \,,
\end{equation}
for $0 < \tau \leq \tau_d(s_d,s_b,\alpha)$,
and the boundary condition for newborn cells
\begin{equation}
	\label{eq_bc_fw}
	p(\tau=0,s_d,s_b,\alpha;t) = \int \di s_d' \di s_b' \di \alpha' \ 
	\mathcal{K}(s_d,s_b,\alpha|s_d',s_b',\alpha') p(\tau_d',s_d',s_b',\alpha';t) \,.
\end{equation}

This probability distribution can also be recovered from population data by a non-uniform sampling of the lineages of the population tree, called the forward sampling \cite{nozoe_inferring_2017}. In the following we adopt this terminology, widely used in the literature, and call \textit{forward lineages} these single lineages.

For forward lineages, there is no distinction between distributions including and excluding leaves, since there is only one leaf at the end of a lineage, but an infinity of non-leaf cells along the lineage in the steady-state limit. Therefore, the unique steady-state joint probability density along a forward lineage is defined as 
\begin{equation}
	\rho_{\mathrm{fw}}(s_d,s_b,\alpha) = \frac{p(0,s_d,s_b,\alpha)}{\int \di s_d' \di s_b' \di \alpha'  \ p(0,s_d',s_b',\alpha')} \,.
\end{equation}

\section{Derivation of the theoretical results}

\subsection{General lineage-population bias (eq. 12 in the main text)}
\label{app_powell}

In this section, we derive the lineage-population bias for the division size distribution conditioned on birth size and single-cell growth rate, eq. 12 in the main text.

At the forward level, the steady-state solution to \cref{eq_pde_fw} is
\begin{equation}
	p(\tau,s_d,s_b,\alpha) = \begin{cases}
		p(0 ,s_d,s_b,\alpha) & \tau \leq \tau_d \\
		0 & \tau > \tau_d \,,
	\end{cases}
\end{equation}
which confirms that the distributions with and without the leaf cell are identical in the forward statistics, as explained in the previous section, since $p(\tau_d,s_d,s_b,\alpha)=p(0,s_d,s_b,\alpha)$. 
At the population level, the steady-state solution to \cref{eq:pbe_alpha} reads
\begin{equation}
	\label{eq_Pi_sol}
	\Pi(\tau,s_d,s_b,\alpha) = \begin{cases}
		\Pi(0 ,s_d,s_b,\alpha) e^{-\lambda \tau} & \tau \leq \tau_d \\
		0 & \tau > \tau_d \,.
	\end{cases}
\end{equation}

We now recast the boundary conditions \cref{eq_bc_fw,eq:bc_alpha} in terms of the normalized distributions:
\begin{align}
	\label{eq_BC_rho_fw}
	\rho_{\mathrm{fw}}(s_d,s_b,\alpha) &= \int \di s_d' \di s_b' \di \alpha' \ 
	\mathcal{K}(s_d,s_b,\alpha|s_d',s_b',\alpha') \rho_{\mathrm{fw}}(s_d',s_b',\alpha') \\
	\label{eq_BC_rho_nb}
	\rho_{\mathrm{tree}}^{\mathrm{nb}}(s_d,s_b,\alpha) &= 2 \int \di s_d' \di s_b' \di \alpha' \ 
	\mathcal{K}(s_d,s_b,\alpha|s_d',s_b',\alpha') \rho_{\mathrm{tree}}^{\mathrm{nb}}(s_d',s_b',\alpha') e^{-\lambda \tau_d'}\,.
\end{align}
From \cref{eq_BC_rho_fw,eq_BC_rho_nb}, we see that the decomposition of division kernel $\mathcal{K}$, given in \cref{eq_K_kernel}, imposes 
\begin{equation}
	\rho_{\mathrm{fw}}(s_d|s_b,\alpha)=\rho_{\mathrm{tree}}^{\mathrm{nb}}(s_d|s_b,\alpha)=\psi(s_d|s_b,\alpha) \,.
\end{equation}
This means that there is no bias in the cell size control mechanism when comparing a forward lineage and a population tree with its leaves. However, the division sizes of the leaves are inaccessible by definition. Excluding the leaves from the tree statistics is what creates a bias, as:
\begin{align}
	\rho_{\mathrm{tree}}(s_d,s_b,\alpha) &=  \frac{ \Pi(\tau_d,s_d,s_b,\alpha)}{\int \di s_d' \di s_b' \di \alpha' \ \Pi(\tau_d',s_d',s_b',\alpha')} \\   
	&=  \frac{ \Pi(0,s_d,s_b,\alpha)e^{-\lambda \tau_d}}{\int \di s_d' \di s_b' \di \alpha' \ \Pi(\tau_d',s_d',s_b',\alpha')} \\  
	&= 2 \rho_{\mathrm{tree}}^{\mathrm{nb}}(s_d,s_b,\alpha) e^{-\lambda \tau_d} \\
	\label{eq_gen_powell}
	&= 2 \rho_{\mathrm{tree}}^{\mathrm{nb}}(s_b,\alpha) \psi(s_d|s_b,\alpha) e^{-\lambda \tau_d} \,.
\end{align}
From here, we derive two results: the bias between the newborn and tree distributions of birth sizes, and the bias between the population and forward distributions of division size conditioned on birth size and single cell growth rate, which defines the mechanism of cell size control. 

Conditioning \cref{eq_gen_powell} on the single cell growth rate $\alpha$ and integrating over the division size $s_d$ leads to
\begin{equation}
	\label{rho_sb_bias}
	\rho_{\mathrm{tree}}(s_b|\alpha) = \frac{2  \rho_{\mathrm{tree}}^{\mathrm{nb}}(s_b|\alpha) \int \di s_d \ e^{-\lambda \tau_d(s_d,s_b,\alpha)} \psi(s_d|s_b,\alpha)}{Z(\alpha)} \,,
\end{equation}
where $Z(\alpha)=\rho_{\mathrm{tree}}(\alpha)/\rho_{\mathrm{tree}}^{\mathrm{nb}}(\alpha)$ is a $s_b$-independent normalizing factor. This provides the bias between the newborn and tree distributions of conditional birth sizes.

Conditioning \cref{eq_gen_powell} on $s_b$ and $\alpha$ gives eq. 12 in the main text:
\begin{equation}
	\label{app_powell_cond_sd}
	\rho_{\mathrm{tree}}(s_d|s_b,\alpha) = \frac{2  \rho_{\mathrm{fw}}(s_d|s_b,\alpha) e^{-\lambda \tau_d(s_d,s_b,\alpha)}}{Z(s_b,\alpha)} \,,
\end{equation}
with $Z(s_b,\alpha)=\rho_{\mathrm{tree}}(s_b,\alpha)/\rho_{\mathrm{tree}}^{\mathrm{nb}}(s_b,\alpha)$ a $s_d$-independent normalization factor.

Note that a change of variable from the division size $s_d$ to the generation time $\tau_d$, $\phi(\tau_d|s_b,\alpha) d \tau_d=\rho(s_d|s_b,\alpha) d s_d $, in \cref{app_powell_cond_sd} gives
\begin{equation}
	\label{powell_cond}
	\phi_{\mathrm{tree}}(\tau_d|s_b,\alpha)= \frac{ 2 \phi_{\mathrm{fw}}(\tau_d|s_b,\alpha) e^{-\lambda \tau_d}}{Z(s_b,\alpha)}  \,,
\end{equation}
which is a generalization of Powell's bias on generation times between the lineage and population levels, initially derived for age-controlled cells \cite{powell_growth_1956}, to arbitrary size control mechanism.

\subsection{General properties of the lineage-population bias}

\subsubsection{General inequalities}
\label{sec_ineq}

In this section, we derive a series of inequalities that are valid beyond the small noise assumption presented in the main text.

Generation time $\tau_d(s_d,s_b,\alpha)$ is an increasing function of division size $s_d$, \cref{app_powell_cond_sd} therefore implies
\begin{equation}
	\label{eq_ineq_sd}
	\langle s_d | s_b,\alpha \rangle_{\mathrm{tree}} \leq \langle s_d | s_b,\alpha \rangle_{\mathrm{fw}} \,.
\end{equation}
Subtracting $s_b$ to both sides gives
\begin{equation}
	\label{eq_ineq_delta}
	\langle \Delta_d | s_b,\alpha \rangle_{\mathrm{tree}} \leq \langle \Delta_d | s_b,\alpha \rangle_{\mathrm{fw}} \,.
\end{equation} 
Cells born with the same size and growth rate divide, on average, after growing by a size increment that is smaller in populations than in forward lineages.
This further implies an unconditioned inequality for the adder. Indeed, if the mechanism is an adder in either one of the two statistics, i.e.
$\langle \Delta_d|s_b,\alpha \rangle_{\mathrm{fw}}=\langle \Delta_d |\alpha\rangle_{\mathrm{fw}}$
or $\langle \Delta_d|s_b,\alpha \rangle_{\mathrm{tree}}=\langle \Delta_d |\alpha\rangle_{\mathrm{tree}}$,
then integrating out $s_b$ in \cref{eq_ineq_delta} shows that, on average, cells grow by a smaller added size before dividing in populations than in forward lineages:
\begin{equation}
	\label{eq_ineq_Del_add}
	\langle \Delta_d|\alpha \rangle_{\mathrm{tree}} \leq \langle \Delta_d|\alpha \rangle_{\mathrm{fw}} \,.
\end{equation}
For the usual adder model, which ignores fluctuations in single-cell growth rate, this inequality reduces to $\langle \Delta_d \rangle_{\mathrm{tree}} \leq \langle \Delta_d \rangle_{\mathrm{fw}}$. Added size can be seen as a physiological age since it increases with time and it is reset to 0 at division \cite{olivier_how_2017}, therefore this inequality can be put in parallel to the one on generation time $\tau_d$ (biological age): $\langle \tau_d \rangle_{\mathrm{tree}} \leq \langle \tau_d \rangle_{\mathrm{fw}}$, that can be derived directly from \cref{powell_cond} \cite{powell_growth_1956}.

\subsubsection{The sizer is conserved at the population level}
\label{sec_spe_cases}

In this section we explain why a forward sizer is conserved at the population level for a broad class of single-cell growth laws.

When the inter-division time can be written as $\tau_d(s_d,s_b,\alpha)=f_1(s_d,\alpha)-f_2(s_b,\alpha)$ for any two functions $f_1$ and $f_2$, then $\rho_{\mathrm{tree}}(s_d|s_b,\alpha)$ in \cref{app_powell_cond_sd} takes the form of a product of a function of $s_d$, $\Tilde{f}_1(s_d,\alpha)=2\rho_{\mathrm{fw}}(s_d|\alpha) e^{-\lambda f_1(s_d,\alpha)}$ and a function of $s_b$, $\Tilde{f}_2(s_b,\alpha)= e^{\lambda f_2(s_b,\alpha)} / Z(s_b,\alpha)$. 
The normalization of the conditional tree distribution reads $\int_{s_b}^{\infty} \di s_d \ \Tilde{f}_1(s_d,\alpha) = 1/\Tilde{f}_2(s_b,\alpha)$. 
In the small noise limit, the distributions of birth sizes and of division sizes have negligible overlap, so we let the lower limit of the integral go to $0$ \cite{taheri-araghi_cell-size_2015,genthon_noisy_2025}, which implies that $\Tilde{f}_2(s_b,\alpha)$ is independent of $s_b$, and thus $\rho_{\mathrm{tree}}(s_d|s_b,\alpha) \equiv \rho_{\mathrm{tree}}(s_d|\alpha)$. 
Therefore the cell size control mechanism remains a sizer at the population level, as observed in 
eqs. 2a and 6a
in the main text for exponentially-growing cells, with non-fluctuating and fluctuating single cell growth rates respectively.

The inter-division time decomposition covers a broad range of single-cell growth laws $ds/d\tau=\alpha s^{\beta}$; which include linear growth $\beta=0$ for which $f_1(x,\alpha)=f_2(x,\alpha)=x/\alpha$, exponential growth $\beta=1$ for which $f_1(x,\alpha)=f_2(x,\alpha)=\ln (x)/\alpha$, and sub- and super-exponential growth, $0<\beta<1$ and $\beta>1$ respectively, for which $f_1(x,\alpha)=f_2(x,\alpha)=x^{1-\beta}/(\alpha(1-\beta))$.

In 
eq. 8a
in the main text, we showed that if fluctuating growth rates are integrated out on the other hand, the sizer is not necessarily conserved at the population level. This is because
$\rho_{\mathrm{tree}}(s_d|s_b)= \int \di \alpha \ \rho_{\mathrm{tree}}(s_d|s_b,\alpha)\rho_{\mathrm{tree}}(\alpha|s_b) \equiv \int \di \alpha \ \rho_{\mathrm{tree}}(s_d|\alpha)\rho_{\mathrm{tree}}(\alpha|s_b)$, where $\rho_{\mathrm{tree}}(\alpha|s_b)$ carries a dependence on $s_b$ since $\alpha$ and $s_b$ are not independent in the tree distribution without leaves.

\subsection{Constant growth rates}

We derive here the results from the first part of the main text, where single-cell growth rates are non-fluctuating, and therefore $\lambda=\alpha$.

\subsubsection{Lineage-population bias for mean birth size (eq. 3 in the main text)}
\label{sec_bias_sb}

For exponentially growing cells, the bias between the birth size distribution in the tree ensemble with and without leaves, \cref{rho_sb_bias}, takes the simple form:
\begin{equation}
	\label{dis_bias_sb_div_nb}
	\rho_{\mathrm{tree}}(s_b) = 2  s_b  \langle s_d^{-1}|s_b \rangle_{\mathrm{fw}} \rho_{\mathrm{tree}}^{\mathrm{nb}}(s_b) \,.
\end{equation}
The mean birth size is thus given by 
\begin{equation}
	\label{eq_sb_mean_step}
	\langle s_b \rangle_{\mathrm{tree}} = \frac{\langle s_b^2  \langle s_d^{-1}|s_b \rangle_{\mathrm{fw}} \rangle_\mathrm{tree}^\mathrm{nb}}{\langle s_b  \langle s_d^{-1}|s_b \rangle_{\mathrm{fw}} \rangle_\mathrm{tree}^\mathrm{nb}} \,,
\end{equation}
where it is convenient to express the factor $2$ in \cref{dis_bias_sb_div_nb} as a normalization.
We first Taylor expand the division size $s_d$, given by the linear map $s_d=a s_b + b + \eta$, around its forward mean value $\langle s_d | s_b \rangle_\mathrm{fw} =a s_b +b$: 
\begin{equation}
	s_d^{-1} = \frac{1}{a s_b + b} \sum_{n=0}^{\infty} \frac{(-\eta)^n}{(a s_b + b)^n} \,.
\end{equation} 
Since $\eta$ is a zero-mean noise, upon taking the forward average we have
\begin{equation}
	\label{eq_sd_inv_series}
	\langle s_d^{-1} | s_b \rangle_\mathrm{fw} = \frac{1}{a s_b + b} \lbk 1 + \sum_{n=2}^{\infty} \frac{(-1)^n \langle \eta^n | s_b \rangle_\mathrm{fw}}{(a s_b + b)^n} \rbk \,. 
\end{equation}
We seek leading order correction for $\langle s_b \rangle_{\mathrm{tree}}$ in the small noise limit, and since $\langle s_d^{-1} | s_b \rangle_\mathrm{fw}$ appears on both the numerator and denominator of \cref{eq_sb_mean_step}, moments of noise $\eta$ will actually not play a role. Indeed, let us keep only the leading order correction in the variance of the noise on division size in \cref{eq_sd_inv_series}:
\begin{equation}
	\label{eq_in_sd_fw}
	\langle s_d^{-1}|s_b \rangle_{\mathrm{fw}} \approx \frac{1}{a s_b + b} \lbk 1 + \frac{\langle \eta^2 | s_b \rangle_\mathrm{fw}}{(a s_b+b)^2} \rbk \,,
\end{equation} 
expand the birth size $s_b$ around its mean value $\langle s_b \rangle_{\mathrm{tree}}^{\mathrm{nb}}$ in the numerator and denominator of \cref{eq_sb_mean_step}, and keep only leading order terms (i.e. getting rid of corrections of the order of $\sigma^2[s_b]\langle \eta^2 | s_b \rangle $):
\begin{equation}
	\label{eq_bias_sb_div_nb}
	\langle s_b \rangle_{\mathrm{tree}} \approx \langle s_b \rangle_{\mathrm{tree}}^{\mathrm{nb}} \lbk 1 + \frac{b}{a \langle s_b \rangle_{\mathrm{tree}}^{\mathrm{nb}} + b} \mathrm{CV}_{\mathrm{tree, nb}}^2[s_b] \rbk \,.
\end{equation}
This result shows that including the leaf cells in the tree statistics always decreases the average birth size

To compute the birth size statistics in the tree ensemble with leaf cells, we use a method developed in \cite{thomas_analysis_2018} (in this reference, the "tree" distribution includes leaf cells and correspond to our distribution $\rho_{\mathrm{tree}}^{\mathrm{nb}}$).

First, using the division kernel specified by \cref{eq_K_kernel,eq:def_kappa}, we integrate out $s_d$ in the boundary conditions for newborn cells in the forward and tree statistics with leaves, \cref{eq_BC_rho_fw} and \cref{eq_BC_rho_nb}, which after some re-arrangements gives:
\begin{align}
	\label{eq_fw_rec_rel}
	\rho_{\mathrm{fw}}(s_b) &= \int \di p \di s_b'  \ 
	\rho_{\mathrm{fw}}(s_b')\rho_{\mathrm{fw}}(s_b/p|s_b') \frac{\pi(p)}{p} \\
	\label{eq_tree_rec_rel}
	\rho_{\mathrm{tree}}^{\mathrm{nb}}(s_b) &= 2 \int \di p \di s_b' \ 
	\rho_{\mathrm{tree}}^{\mathrm{nb}}(s_b') \rho_{\mathrm{fw}}(s_b/p|s_b') \frac{s_b'}{s_b} \pi(p)\,.
\end{align}
All the forward moments can be computed using this recursion relation \cite{thomas_analysis_2018}. Here, we focus on the mean and the variance. We multiply \cref{eq_fw_rec_rel} by $s_b$ and integrate, giving for the mean birth size:
\begin{equation}
	\label{eq_sb_fw_mean_p}
	\langle s_b \rangle_\mathrm{fw} = \frac{b \langle p \rangle}{1 - a \langle p \rangle} \,.
\end{equation}
For binary fission, $\langle p \rangle=1/2$, we recover the general result $ \langle s_b \rangle_\mathrm{fw}=b/(2-a)$.
Following similar steps for the second moment after multiplying \cref{eq_fw_rec_rel} by $s_b^2$, we obtain at leading order for each type of noise:
\begin{equation}
	\label{eq_CV_fw}
	\mathrm{CV}^2_\mathrm{fw}[s_b] \approx \frac{4}{4-a^2} \mathrm{CV}_p^2 + \frac{2-a}{2+a} \frac{\sigma_a^2}{b^2} + \frac{2}{2+a}\frac{\sigma_i^2}{b} + \frac{4}{4-a^2} \sigma_e^2 \,.
\end{equation}

Second, we identify in \cref{eq_tree_rec_rel} that the distribution along backward lineages \cite{thomas_analysis_2018}
\begin{equation}
	\label{eq_def_rho_bw}
	\rho_{\mathrm{bw}}(s_b) = s_b \rho_{\mathrm{tree}}^{\mathrm{nb}}(s_b) \langle s_b^{-1} \rangle_\mathrm{bw}
\end{equation}
follows the equation
\begin{equation}
	\rho_{\mathrm{bw}}(s_b) = \int \di p \di s_b'  \ 
	\rho_{\mathrm{bw}}(s_b')\rho_{\mathrm{fw}}(s_b/p|s_b') \frac{\pi_\mathrm{bw}(p)}{p}
\end{equation}
which has the same form as \cref{eq_fw_rec_rel} for forward lineages, but with modified volume partitioning at division:
\begin{equation}
	\label{eq_pi_bias}
	\pi_\mathrm{bw}(p) = 2 p \pi(p) \,.
\end{equation}
Therefore, the backward moments are obtained from the forward moments by replacing $\pi(p)$ by $2p\pi(p)$. The backward mean is obtained by replacing $\langle p \rangle$ by $2\langle p^2 \rangle$ in \cref{eq_sb_fw_mean_p}, with $\langle p^2 \rangle=(1+\mathrm{CV}_p^2)/4$:
\begin{equation}
	\label{eq_sb_bw_mean}
	\langle s_b \rangle_\mathrm{bw} = \frac{b (1+\mathrm{CV}_p^2)}{2 - a (1+\mathrm{CV}_p^2)} \,.
\end{equation}
The backward CV is obtained in the same way, and is equal to the forward CV at leading order: $\mathrm{CV}^2_\mathrm{bw}[s_b] \approx \mathrm{CV}^2_\mathrm{fw}[s_b]$.

Third, \cref{eq_def_rho_bw} implies $\langle s_b \rangle_\mathrm{tree}^\mathrm{nb} = 1/\langle s_b^{-1} \rangle_\mathrm{bw}$. A small noise expansion of birth size around its backward mean at leading order reads $\langle s_b \rangle_\mathrm{tree}^\mathrm{nb} \approx \langle s_b \rangle_\mathrm{bw} (1-\mathrm{CV}_\mathrm{fw}^2[s_b])$.
Combining this expression with \cref{eq_sb_fw_mean_p,eq_CV_fw,eq_sb_bw_mean} gives the mean birth size in the tree statistics with leaves at leading order:
\begin{equation}
	\label{eq_sb_tree_mean}
	\langle s_b \rangle_{\mathrm{tree}}^\mathrm{nb} \approx \langle s_b \rangle_{\mathrm{fw}} \lbk  1 - \frac{2-a}{(2+a)b^2}\sigma_a^2 - \frac{2}{(2+a)b}\sigma_i^2 - \frac{4}{4-a^2}\sigma_e^2 + \frac{2a}{4-a^2} \mathrm{CV}_p^2 \rbk \,.
\end{equation}
Again, the CV is unbiased at leading order: $\mathrm{CV}_{\mathrm{tree, nb}}[s_b] \approx \mathrm{CV}_{\mathrm{fw}}[s_b]$.

Finally, plugging \cref{eq_sb_tree_mean,eq_CV_fw} in \cref{eq_bias_sb_div_nb}, gives at leading order
\begin{equation}
	\label{eq_sb_tree_div_SM}
	\langle s_b \rangle_{\mathrm{tree}} \approx \langle s_b \rangle_{\mathrm{fw}} \lbk  1 - \frac{a(2-a)}{2(2+a)b^2}\sigma_a^2 - \frac{a}{(2+a)b}\sigma_i^2 - \frac{2a}{4-a^2}\sigma_e^2 + \frac{4}{4-a^2} \mathrm{CV}_p^2 \rbk \,,
\end{equation}
which is eq. 3 form the main text.

\begin{figure}
	\centering
	\includegraphics[width=\linewidth]{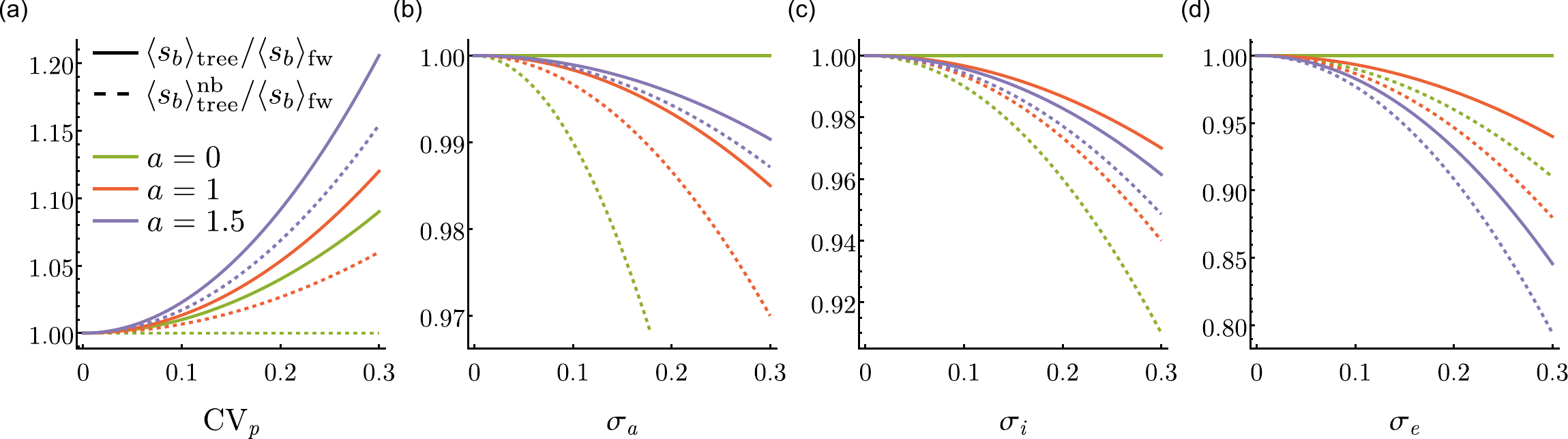}
	\caption{Noises on division size and on division asymmetry modulate mean birth sizes in the tree statistics both with and without leaf cells.}
	\label{fig:sb_mean_vs_leaves}
\end{figure}

We illustrate the impact of the different sources of noise on the mean birth size both with and without leaves in \cref{fig:sb_mean_vs_leaves}. 

\subsubsection{Population-level noisy linear map (eq. 2 in the main text)}
\label{sec_pop_map}

\begin{figure}
	\centering
	\includegraphics[width=0.8\linewidth]{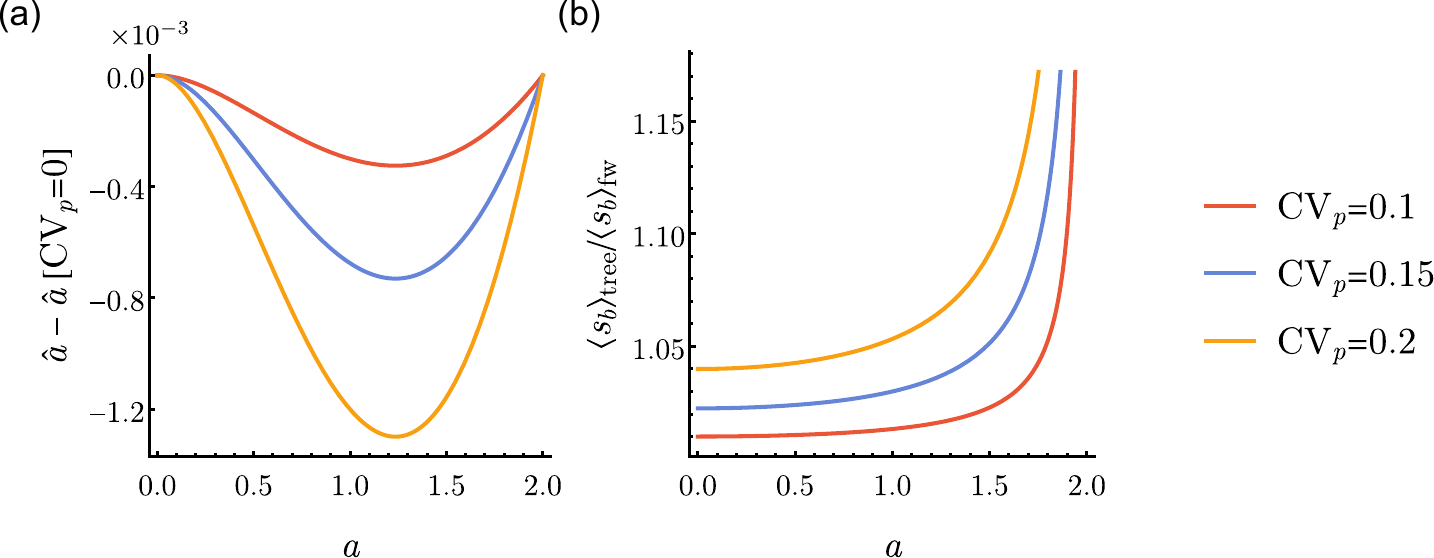}
	\caption{Impact of noise in division asymmetry (a) on population slope modulation in presence of additive noise with $\sigma_a=0.3$, and (b) on population average birth size.}
	\label{fig:CVp}
\end{figure}

In this section, we detail the derivation of eq. 2 from the main text, sketched in the Methods section, and we discuss higher order corrections from intrinsic noise and noise in division asymmetry. 

We start from 
eq. 13
from the main text:
\begin{equation}
	\rho_{\mathrm{tree}}(s_d|s_b) = \frac{s_d^{-1}  \rho_{\mathrm{fw}}(s_d|s_b) }{\langle s_d^{-1} | s_b \rangle_\mathrm{fw}} \,.
\end{equation}
The conditional mean division size is therefore given by 
\begin{equation}
	\langle s_d |s_b \rangle_\mathrm{tree} = \frac{1}{\langle s_d^{-1} | s_b \rangle_\mathrm{fw}} \,.
\end{equation}
Using the expansion of the division size in powers of noise $\eta$, \cref{eq_sd_inv_series}, in the small noise limit we only keep the terms $n=2$ and $n=3$ in the series:
\begin{equation}
	\label{eq_sd_inv_n3}
	\langle s_d^{-1} | s_b \rangle_\mathrm{fw} \approx \frac{1}{a s_b + b} \lbk 1 + \frac{ \langle \eta^2 | s_b \rangle_\mathrm{fw}}{(a s_b + b)^2} - \frac{ \langle \eta^3 | s_b \rangle_\mathrm{fw}}{(a s_b + b)^3} \rbk \,. 
\end{equation}
Then, we consider that noise on division size is a combination of independent additive, intrinsic and extrinsic noises, $\eta=\eta_a + \sqrt{a s_b + b} \eta_i + (a s_b +b) \eta_e$, which leads to:
\begin{equation}
	\label{eq_sd_exp_3}
	\langle s_d | s_b \rangle_{\mathrm{tree}} \approx (as_b+b) \lbk 1 -  \frac{\sigma_a^2}{(as_b +b)^2} - \frac{\sigma_i^2}{as_b +b} - \sigma_e^2 
	+ \frac{\gamma_a \sigma_a^3}{(as_b +b)^3} + \frac{\gamma_i \sigma_i^3}{(as_b +b)^{3/2}} + \gamma_e \sigma_e^3 
	\rbk\,,
\end{equation}
with $\gamma_k=\langle \eta_k^3 \rangle / \sigma_k^3$ the skewness of noise $\eta_k$. We explained in the main text that at leading order, intrinsic noise only modulates the intercept of the linear map but not the slope. We observe here that the next order correction in $\sigma_i^3$ is a non-linear function of birth size $s_b$, and will thus modulate both the slope and intercept once linearized.
We now linearize \cref{eq_sd_exp_3} around the mean birth $\langle s_b \rangle_\mathrm{tree}$, keeping leading order corrections for each type of noise (i.e. neglecting terms in $\sigma_a^3$ and $\sigma_e^3$), which gives the deterministic part of the tilted linear map at the population level: $s_d = \hat{a} s_b + \hat{b}$, with modified slope and intercept:
\begin{align}
	\label{eq_hat_a_inter}
	\hat{a}&\approx a \lbk 1 +  \frac{b^2}{(a\langle s_b \rangle_\mathrm{tree}+b)^2} \frac{\sigma_a^2}{b^2} 
	- \sigma_e^2
	-  \frac{\gamma_i}{2} \frac{\sigma_i^3}{(a\langle s_b \rangle_\mathrm{tree}+b)^{3/2}}  \rbk \\
	\label{eq_hat_b_inter}
	\hat{b}&\approx b \lbk 1 - \frac{b(2a \langle s_b \rangle_\mathrm{tree} +b)}{(a \langle s_b \rangle_\mathrm{tree} +b)^2} \frac{\sigma_a^2}{b^2} - \frac{\sigma_i^2}{b} - \sigma_e^2 + \frac{\gamma_i (3a \langle s_b \rangle_\mathrm{tree} +2b)}{2(a \langle s_b \rangle_\mathrm{tree} +b)^{3/2}} \sigma_i^3 \rbk \,.
\end{align}

Finally, we replace the mean birth size by its expression given in \cref{eq_sb_tree_div_SM} and we keep leading order corrections for each source of noise:
\begin{align}
	\label{eq_a_pop_lin_map}
	\hat{a} &\approx a \lbk 1 + \frac{(2-a)^2}{4b^2} \sigma_a^2 - \sigma_e^2 - \frac{(2-a)^{3/2}}{4 \sqrt{2} b^{3/2}} \gamma_i\sigma_i^3\rbk - \frac{(2-a)a^2}{(2+a)b^2}  \lbk \sigma_a^2 - \frac{3\sqrt{b}}{4 \sqrt{2(2-a)}}\gamma_i\sigma_i^3 \rbk \mathrm{CV}_p^2
	\\
	\label{eq_b_pop_lin_map}
	\hat{b} &\approx b \lbk 1 - \frac{4-a^2}{4b^2} \sigma_a^2 -\frac{1}{b}\sigma_i^2 - \sigma_e^2  \rbk +\frac{a^2}{(2+a)b}\lbk \sigma_a^2 - \frac{3\sqrt{b}}{4\sqrt{2(2-a)}} \gamma_i\sigma_i^3 \rbk \mathrm{CV}_p^2
\end{align}
Noise in division asymmetry never modulates the linear map when coupled to extrinsic noise only because corrections in $\mathrm{CV}_p^2$ come from the linearization around the mean birth size in the tree statistics and no linearization is necessary since extrinsic noise conserves the linear structure of the forward map. 
On the other hand, noise in division asymmetry coupled to additive noise decreases the slope, as shown in \cref{fig:CVp}a, and increases the intercept, thus acting against the modulation of the linear map due to additive noise alone. 
Finally, we observe that noise in division asymmetry modulates the slope and the intercept when coupled to intrinsic only at third order in $\sigma_i$. This is because at second order, intrinsic noise preserves the linear structure of the forward linear map.

Finally, we compute the second moment of the conditional division size distribution:
\begin{align}
	\langle s_d^2 |s_b \rangle_\mathrm{tree} &= \frac{\langle s_d | s_b \rangle_\mathrm{fw}}{\langle s_d^{-1} | s_b \rangle_\mathrm{fw}} \\
	&=(a s_b + b) \langle s_d |s_b \rangle_\mathrm{tree} \,.
\end{align}
Using \cref{eq_sd_exp_3}, the variance of the population linear map is given by:
\begin{align}
	\langle \hat{\eta}^2|s_b\rangle_{\mathrm{tree}} &\approx\sigma_a^2 \lp 1 -\frac{\gamma_a \sigma_a}{a s_b +b} \rp + (a s_b+b) \sigma_i^2 \lp 1 -\frac{ \gamma_i \sigma_i}{\sqrt{a s_b +b}} \rp + (a s_b+b)^2 \sigma_e^2 \lp 1 -\gamma_e \sigma_e \rp \,.
\end{align}

\subsubsection{Inequality on added size}

The average added size in the tree statistics $\langle \Delta_d \rangle_{\mathrm{tree}}=(\hat{a}-1)\langle s_b \rangle_{\mathrm{tree}} +\hat{b}$, where $\hat{a}$ and $\hat{b}$ are given in \cref{eq_a_pop_lin_map,eq_b_pop_lin_map} and $\langle s_b \rangle_{\mathrm{tree}}$ is given in \cref{eq_sb_tree_div_SM}, is biased compared to the forward average at leading order by: 
\begin{equation}
	\langle \Delta_d \rangle_{\mathrm{tree}}  \approx \langle \Delta_d \rangle_{\mathrm{fw}} \lbk 1 
	- \frac{4(1-a)}{4-a^2} \mathrm{CV}_{p}^2 
	- \frac{(4-a)(2-a)}{2(2+a)}\frac{\sigma_a^2}{b^2} 
	- \frac{4-a}{2+a} \frac{\sigma_i^2}{b}
	- \frac{2(4-a)}{4-a^2} \sigma_e^2 \rbk \,.
\end{equation}
Additive, intrinsic and extrinsic noises on division size decrease the average added size, while the impact of noise in division asymmetry depends on the sign of $1-a$. When $a \leq 1$, the average added size is smaller in the tree statistics than in the forward statistics, which generalizes the inequality derived in \cref{sec_ineq} for the adder only.

\subsubsection{Data collapse on rescaled extrinsic noise model (fig. 3a-c)}

We explain here under which conditions data collapse on the master curve $\sigma_{\mathrm{fw}}^2[s_d|s_b]/\sigma_{\mathrm{fw}}^2[s_d]= 0.75 \langle s_d | s_b \rangle_\mathrm{fw}^2/\langle s_d \rangle_\mathrm{fw}^2$, as shown in fig. 3a-c in the main text. 

The extrinsic noise model is defined by
\begin{equation}
	\sigma_{\mathrm{fw}}^2[s_d|s_b]= \sigma_e^2 \langle s_d | s_b \rangle_\mathrm{fw}^2 \,,
\end{equation}
where the extrinsic variance $\sigma_e^2$ contributes to the total noise in division size. There are several routes to derive $\mathrm{CV}^2_\mathrm{fw}[s_d]$, and the most direct one is to link it to the total noise in birth sizes, $\mathrm{CV}^2_\mathrm{fw}[s_b]$, which we already computed. At division, a daughter cell inherits a fraction $p$ of its mother size: $s_b = p s_d$. Since $p$ and $s_d$ are independent random variables, in the small noise limit we have $\mathrm{CV}^2_\mathrm{fw}[s_b] \approx \mathrm{CV}_{p}^2 + \mathrm{CV}^2_\mathrm{fw}[s_d]$. Therefore, using \cref{eq_CV_fw} with extrinsic noise ($\sigma_a=0$ and $\sigma_i=0$), we obtain:
\begin{equation}
	\mathrm{CV}^2_{\mathrm{fw}}[s_d] \approx \frac{a^2}{4-a^2}\mathrm{CV}_{p}^2 + \frac{4}{4-a^2}\sigma_e^2 \,.
\end{equation}
Combining these two relations gives:
\begin{equation}
	\frac{\sigma_{\mathrm{fw}}^2[s_d|s_b]}{\sigma_{\mathrm{fw}}^2[s_d]} \approx \lp \frac{4-a^2}{4} - \frac{a^2}{4} \frac{\mathrm{CV}_{p}^2}{\mathrm{CV}^2_{\mathrm{fw}}[s_d]}\rp \frac{\langle s_d | s_b \rangle_\mathrm{fw}^2}{\langle s_d \rangle_\mathrm{fw}^2} \,.
\end{equation}
We recover the factor $0.75$ for data following an adder $a=1$ with negligible noise in division asymmetry $\mathrm{CV}_{p}^2 \approx 0$.

\subsection{Fluctuating growth rates}

In this section, we derive the results from the second part of the main text when single-cell growth rates are fluctuating, with extrinsic noise and growth-rate-dependent size control as defined by eq. 4 in the main text, reminded here: 
\begin{align}
	\label{eq_SM_a_alp}
	a(\alpha) &= a \lbk 1  + S_a \frac{\alpha - \langle \alpha \rangle_\mathrm{fw}}{\langle \alpha \rangle_\mathrm{fw}} \rbk \\
	\label{eq_SM_b_alp}
	b(\alpha) &= b \lbk 1  + S_b \frac{\alpha - \langle \alpha \rangle_\mathrm{fw}}{\langle \alpha \rangle_\mathrm{fw}} \rbk \\
	\label{eq_SM_s_alp}
	\sigma^2_e(\alpha) &= \sigma_e^2 \lbk 1  + S_{\sigma} \frac{\alpha - \langle \alpha \rangle_\mathrm{fw}}{\langle \alpha \rangle_\mathrm{fw}} \rbk \,.
\end{align}
For more compact notations, we also define (eq. 5 in the main text):
\begin{equation}
	S_\Delta = a S_a + (2-a) S_b \,.
\end{equation}

\subsubsection{Preliminary relations (including eq. 10 in the main text)}

We start by deriving a series of results which will be useful in the next sections.

First, the factorized form of division kernel $\mathcal{K}$, given in \cref{eq_K_kernel}, implies that birth size and single-cell growth rate are independent random variables in the forward and the newborn statistics (see \cref{eq_BC_rho_fw,eq_BC_rho_nb}): $\rho_\mathrm{fw}(s_b,\alpha)=\rho_\mathrm{fw}(s_b)\rho_\mathrm{fw}(\alpha)$ and $\rho_\mathrm{tree}^\mathrm{nb}(s_b,\alpha)=\rho_\mathrm{tree}^\mathrm{nb}(s_b)\rho_\mathrm{tree}^\mathrm{nb}(\alpha)$, with
\begin{equation}
	\rho_\mathrm{fw}(\alpha)=\rho_\mathrm{tree}^\mathrm{nb}(\alpha)=\nu(\alpha) \,.
\end{equation}

Second, the population growth rate, the mean birth size in the newborn statistics and the CV of birth sizes have been computed in \cite{genthon_noisy_2025} for growth-rate-dependent size control and additive noise (in this reference, the "tree" distribution includes leaf cells and correspond to our distribution $\rho_{\mathrm{tree}}^{\mathrm{nb}}$). Following the same method for extrinsic noise instead of additive noise, we obtain:
\begin{align}
	\label{eq_SM_lambda}
	\frac{\lambda}{\langle \alpha \rangle_\mathrm{fw}} &\approx 1 - \lbk 1 - \frac{\ln2}{2} - \frac{S_\Delta}{2 \ln 2} \rbk \mathrm{CV}_{\alpha}^2 \\
	\label{eq_SM_sb}
	\langle s_b \rangle_{\mathrm{tree}}^\mathrm{nb} &\approx \langle s_b \rangle_{\mathrm{fw}} \lbk  1  - \frac{4}{4-a^2}\sigma_e^2 + \frac{2a}{4-a^2} \mathrm{CV}_p^2 + \frac{S_\Delta((2+a)\ln 2 - S_\Delta)}{4-a^2} \mathrm{CV}_{\alpha}^2\rbk \\
	\label{eq_SM_CV}
	\mathrm{CV}_{\mathrm{nb}}^2[s_b] &\approx \frac{4}{4-a^2}\mathrm{CV}_{p}^2 + \frac{4}{4-a^2}\sigma_e^2 + \frac{S_\Delta^2}{4-a^2} \mathrm{CV}_{\alpha}^2 \,.
\end{align}
The population growth rate is unchanged at leading order compared to the one derived in \cite{genthon_noisy_2025} for additive noise, since noise on division size only contributes next order corrections. 
For birth size mean and CV, the contributions coming from noises in division asymmetry and in growth rates are the same as in \cite{genthon_noisy_2025}, and the contribution of additive noise (where $\mathrm{CV}_\varphi=\sigma_a/b$ in \cite{genthon_noisy_2025}) has been replaced by that of extrinsic noise.

Third, we derive the CV of division sizes, eq. 10 in the main text. 
At division, a daughter cell inherits a fraction $p$ of its mother size: $s_b = p s_d$. Since $p$ and $s_d$ are independent random variables, in the small noise limit we have $\mathrm{CV}^2_\mathrm{fw}[s_b] \approx \mathrm{CV}_{p}^2 + \mathrm{CV}^2_\mathrm{fw}[s_d]$, where CVs in the different statistics are equal at leading order: $\mathrm{CV}^2_\mathrm{fw}[s_b] \approx \mathrm{CV}_{\mathrm{nb}}^2[s_b]$. Using \cref{eq_SM_CV}, we obtain:
\begin{equation}
	\mathrm{CV}_{\mathrm{fw}}^2[s_d] \approx \frac{a^2}{4-a^2}\mathrm{CV}_{p}^2 + \frac{4}{4-a^2}\sigma_e^2 + \frac{S_\Delta^2}{4-a^2} \mathrm{CV}_{\alpha}^2 \,.
\end{equation}

Fourth, we derive the mean single-cell growth rate in the tree statistics without leaves. 
Integrating birth and division sizes in \cref{eq_gen_powell}, the tree distribution of growth rates is given by
\begin{equation}
	\rho_{\mathrm{tree}}(\alpha)= 2\int d s_b d s_d \
	\psi(s_d|s_b,\alpha) \rho_ {\mathrm{tree}}^{\mathrm{nb}}(s_b) \nu(\alpha) \lp \frac{s_d}{s_b}\rp^{-\lambda/\alpha} \,.
\end{equation}
We multiply this equation by $\alpha$ and integrate over $\alpha$. In the small noise limit, we expand the birth size, the division size and the single-cell growth rate around their means, and we use \cref{eq_SM_lambda,eq_SM_sb,eq_SM_CV}:
\begin{equation}
	\label{eq_alpha_tree_fw}
	\langle \alpha \rangle_\mathrm{tree} \approx  \langle \alpha \rangle_\mathrm{fw} \lbk 1 + \lp \ln 2 - \frac{S_\Delta}{2} \rp \mathrm{CV}^2_\alpha \rbk \,.
\end{equation}

\subsubsection{Population-level noisy linear map (eq. 6 in the main text)}

To derive the tilted linear map in the tree statistics, we start from \cref{app_powell_cond_sd}, which takes the form
\begin{equation}
	\label{eq_powell_full}
	\rho_{\mathrm{tree}}(s_d|s_b,\alpha) = 
	\frac{s_d^{-\lambda/\alpha} \rho_{\mathrm{fw}}(s_d|s_b,\alpha)}{ \langle s_d^{-\lambda/\alpha}|s_b,\alpha \rangle_{\mathrm{fw}} } \,.
\end{equation}
As explained in the Methods section in the main text, extrinsic noise maintains the forward linear structure between division size and birth size and remains extrinsic at the population level, since
\begin{align}
	\langle s_d^n | s_b, \alpha \rangle_\mathrm{tree} &= \frac{\langle s_d^{n-\lambda/\alpha}|s_b,\alpha \rangle_{\mathrm{fw}}}{\langle s_d^{-\lambda/\alpha}|s_b,\alpha \rangle_{\mathrm{fw}}} \\
	&= \langle s_d | s_b, \alpha \rangle_\mathrm{fw}^n \frac{\langle (1 +\eta_e(\alpha))^{n-\lambda/\alpha}|\alpha \rangle_{\mathrm{fw}}}{\langle (1 +\eta_e(\alpha))^{-\lambda/\alpha}|\alpha \rangle_{\mathrm{fw}}} \,,
\end{align}
for any $n$.
Following the steps detailed in \cref{sec_pop_map} for non-fluctuating growth rates, we obtain the population-level tilted linear map
\begin{equation}
	\label{eq_lin_map_hat}
	s_d = \lp \hat{a}(\alpha)  s_b  + \hat{b}(\alpha) \rp \lp 1+ \hat{\eta}_e(\alpha) \rp \,,
\end{equation}
with modified parameters and noise term given by:
\begin{align}
	\hat{a}(\alpha)&\approx a(\alpha) \lbk 1-\frac{\lambda}{\alpha}\sigma_e^2(\alpha) \rbk \\
	\hat{b}(\alpha)&\approx b(\alpha) \lbk 1-\frac{\lambda}{\alpha}\sigma_e^2(\alpha) \rbk \\
	\langle \hat{\eta}_e^2(\alpha) | s_b,\alpha\rangle_{\mathrm{tree}} &\approx \sigma_e^2(\alpha) \lbk 1-\frac{\lambda}{\alpha}\gamma_e\sigma_e(\alpha) \rbk \,,
\end{align}
where the $\alpha$-dependence of the skewness $\gamma_e$ of noise $\eta_e(\alpha)$ is ignored here for simplicity. Note that, because extrinsic noise preserves the linearity between birth and division sizes, no linearization of birth size around its mean is necessary here.
The extrinsic corrections are similar to those derived in \cref{sec_pop_map}, but with a prefactor $\lambda/\alpha$.

We finally linearize these expressions around the mean single-cell growth rate in the tree statistics given by \cref{eq_alpha_tree_fw}. Using the forward sensitivities defined in \cref{eq_SM_a_alp,eq_SM_b_alp,eq_SM_s_alp} and the expression of the population growth rate given in \cref{eq_SM_lambda}, we obtain:
\begin{align}
	\hat{a}(\alpha)&\approx a \lbk 1-\sigma_e^2 +\lp \ln 2 - \frac{S_\Delta}{2}\rp S_a \mathrm{CV}_{\alpha}^2 + \hat{S}_a \frac{\alpha - \langle \alpha \rangle_{\mathrm{tree}}}{\langle \alpha \rangle_{\mathrm{tree}}} \rbk \\
	\hat{b}(\alpha)&\approx b \lbk 1-\sigma_e^2 + \lp \ln 2 - \frac{S_\Delta}{2}\rp S_b \mathrm{CV}_{\alpha}^2 + \hat{S}_b \frac{\alpha - \langle \alpha \rangle_{\mathrm{tree}}}{\langle \alpha \rangle_{\mathrm{tree}}} \rbk \\
	\langle \hat{\eta}_e^2(\alpha) | s_b,\alpha\rangle_{\mathrm{tree}} &\approx \sigma_e^2 \lbk 1-\gamma_e\sigma_e + \lp \ln 2 - \frac{S_\Delta}{2}\rp S_{\sigma} \mathrm{CV}_{\alpha}^2 + \hat{S}_\sigma \frac{\alpha - \langle \alpha \rangle_{\mathrm{tree}}}{\langle \alpha \rangle_{\mathrm{tree}}} \rbk \,,
\end{align}
with sensitivities
\begin{align}
	\hat{S}_a &= S_a \lbk 1+ \lp \ln 2 - \frac{S_\Delta}{2}\rp \mathrm{CV}_{\alpha}^2 \rbk +(1-S_{\sigma}-S_a) \sigma_e^2 \\
	\hat{S}_b &= S_b \lbk 1+ \lp \ln 2 - \frac{S_\Delta}{2}\rp \mathrm{CV}_{\alpha}^2 \rbk +(1-S_{\sigma}-S_b) \sigma_e^2 \\
	\hat{S}_\sigma &=  S_{\sigma} \lbk 1+ \lp \ln 2 - \frac{S_\Delta}{2}\rp \mathrm{CV}_{\alpha}^2 \rbk +\lp 1- \frac{3 S_{\sigma}}{2} \rp \gamma_e\sigma_e \,.
\end{align}

\subsubsection{Added size sensitivity on growth rate at the population level (eq. 7 in the main text)}

The most direct route to derive the population-level sensitivity of added size on growth rate is to start from \cref{eq_gen_powell} conditioned on $\alpha$:
\begin{equation}
	\rho_{\mathrm{tree}}(s_d,s_b|\alpha)  = \
	\frac{(s_d/s_b)^{-\lambda/\alpha}\psi(s_d|s_b,\alpha)\rho_{\mathrm{tree}}^{\mathrm{nb}}(s_b)}{\int \di s_d \di s_b \ (s_d/s_b)^{-\lambda/\alpha}\psi_{\mathrm{fw}}(s_d|s_b,\alpha)\rho_{\mathrm{tree}}^{\mathrm{nb}}(s_b)} \,.
\end{equation}
We then multiply this equation by $\Delta_d=s_d-s_b$, integrate over $s_b$ and $s_d$, expand birth and division sizes around their means using \cref{eq_SM_sb,eq_SM_CV}, and linearize growth rate around its mean value given in \cref{eq_alpha_tree_fw}, which leads to:
\begin{align}
	\langle \Delta_d | \alpha \rangle_{\mathrm{tree}} = & \frac{b}{2-a} \lbk 1  + \frac{4(a-1)}{4-a^2} \mathrm{CV}_p^2 - \frac{2(4-a)}{4-a^2} \sigma_e^2 + \lbk \ln 2 - \frac{S_\Delta}{2} + \frac{(a-1)
		(a S_\Delta -2 \ln 2 (2+a))
	}{2(4-a^2)}\rbk S_\Delta \mathrm{CV}_\alpha^2 + \hat{S}_\Delta \frac{\alpha-\langle\alpha\rangle_{\mathrm{tree}}}{\langle\alpha\rangle_{\mathrm{tree}}} \rbk \\
	\hat{S}_\Delta = & 
	S_\Delta +
	\frac{4 - 2a(3-a)(1-S_a) + a(a-1)S_\Delta}{4-a^2} \mathrm{CV}_p^2 
	+ \frac{4(3-S_\Delta)-a(6+(1-2a)(S_\Delta - 2S_a))}{4-a^2} \sigma_e^2 \\
	&+ \lbk \ln 2 - \frac{S_\Delta}{2} - \frac{2(2 + a(a-3+S_a(1 - 2a)))S_\Delta + (a-1) a S_\Delta^2 + 4 a (2+a) S_a \ln 2}{4(4-a^2)} \rbk S_\Delta \mathrm{CV}_\alpha^2
\end{align}
We recover eq. 7 in the main text for an adder $a=1$.

\subsubsection{Lineage-population bias for mean birth size (for fig. 5j in the main text)}

In fig. 5j in the main text, we show that the prediction from the population tilted linear map agrees with the numerical solution of \cref{app_powell_cond_sd} for the average added size conditioned on birth size and single-cell growth rate. 
In this plot, forward size control is growth-rate-independent ($S_a=S_b=0$) and follows an adder (black line), and $\mathrm{CV}_p=0$.
The population linear map predictions are shown on intervals of $\pm$ 2 standard deviations around the mean birth sizes (shown with circles), where both the mean and the standard deviation depend on single-cell growth rate in the tree statistics. In this section, we show these dependencies for growth-rate-independent forward size control, and for a general mechanism $a$ and non-zero $\mathrm{CV}_p$. 

The bias for the conditional birth size distributions, \cref{rho_sb_bias}, takes the form 
\begin{equation}
	\label{dis_bias_sb_div_nb_alpha}
	\rho_{\mathrm{tree}}(s_b|\alpha) =  \frac{s_b^{\lambda/\alpha}  \langle s_d^{-\lambda/\alpha}|s_b\rangle_{\mathrm{fw}} \rho_{\mathrm{tree}}^{\mathrm{nb}}(s_b)}{\int \di s_b \ s_b^{\lambda/\alpha}  \langle s_d^{-\lambda/\alpha}|s_b \rangle_{\mathrm{fw}} \rho_{\mathrm{tree}}^{\mathrm{nb}}(s_b)} \,.
\end{equation}
Following the same steps as in the absence of fluctuations in single-cell growth rates, detailed in \cref{sec_bias_sb}, we obtain at leading order:
\begin{equation}
	\label{eq_bias_sb_div_nb_alpha}
	\langle s_b | \alpha \rangle_{\mathrm{tree}}  \approx \langle s_b \rangle_{\mathrm{tree}}^{\mathrm{nb}} \lbk 1+ \frac{\lambda}{\alpha} \frac{b}{a \langle s_b \rangle_{\mathrm{tree}}^{\mathrm{nb}} + b } \mathrm{CV}_{\mathrm{tree, nb}}^2[s_b] \rbk \,.
\end{equation}
Expanding \cref{eq_bias_sb_div_nb_alpha} in the small noise limit and using \cref{eq_SM_sb,eq_SM_CV} (with $S_\Delta=0$), we obtain at leading order 
\begin{equation}
	\label{eq_bias_sb_div_fw_cond_alpha_fluct}
	\langle s_b | \alpha \rangle_{\mathrm{tree}}  \approx \langle s_b \rangle_{\mathrm{fw}} \lbk 1+  \frac{2 \lp a+(2-a)\frac{\lambda}{\alpha} \rp}{4-a^2} \mathrm{CV}_{p}^2 - \frac{2 \lp 2-(2-a)\frac{\lambda}{\alpha} \rp}{4-a^2} \sigma_e^2   \rbk \,.
\end{equation}
We recover \cref{eq_sb_tree_div_SM} for non-fluctuating growth rates and extrinsic noise by setting $\lambda=\alpha$.

We derive similarly the second moment and obtain:
\begin{equation}
	\mathrm{CV}^2_\mathrm{tree}[s_b|\alpha] \approx \frac{4}{4-a^2} \mathrm{CV}_p^2 + \frac{4}{4-a^2} \sigma_e^2 + \frac{\lambda}{\alpha} \frac{6 \gamma_\mathrm{fw}[s_b]}{(2+a)\sqrt{4-a^2}} \sigma_e \mathrm{CV}_p^2 + \frac{\lambda}{\alpha} \frac{4 \gamma_\mathrm{fw}[s_b]}{(2+a)\sqrt{4-a^2}} \sigma_e^3 \,.
\end{equation}

\subsubsection{Population linear map when integrating over growth rates (eq. 8 in the main text)}

In typical data analyses, cell size control is inferred by fitting division sizes against birth sizes irrespective of fluctuating growth rates. In this section, we derive the tilted linear map at the population level when growth rates are integrated out. Starting from \cref{eq_gen_powell}, we condition on birth size $s_b$ and integrate over $\alpha$, which gives:
\begin{equation}
	\rho_{\mathrm{tree}}(s_d|s_b)=\frac{\int d\alpha \ \psi(s_d|s_b,\alpha) \nu(\alpha) (s_d/s_b)^{-\lambda/\alpha}}{\int ds_d d\alpha \ \psi(s_d|s_b,\alpha) \nu(\alpha) (s_d/s_b)^{-\lambda/\alpha}} \,.
\end{equation}
Following the same steps as in previous such calculations, we compute the first and second moments of division sizes in the small noise limit by expanding division size and single-cell growth rate in the integrals, and linearizing around the mean birth size. We obtain the population linear map
\begin{equation}
	s_d = \hat{a} s_b + \hat{b} + \hat{\eta} \,,
\end{equation}
with modified slope, intercept and noise term
\begin{align}
	\hat{a}&= a\lbk 1- \sigma_e^2 + \lp \frac{S_\Delta(S_\Delta+2)}{4} + S_a (\ln 2 - S_\Delta)\rp \mathrm{CV}_{\alpha}^2  \rbk - S_\Delta \mathrm{CV}_{\alpha}^2 \\
	\hat{b}&= b\lbk 1- \sigma_e^2 + \lp \frac{S_\Delta(S_\Delta+2)}{4} + S_b (\ln 2 - S_\Delta)\rp \mathrm{CV}_{\alpha}^2  \rbk \\
	\sigma_{\mathrm{tree}}^2[s_d|s_b]&=(a s_b+b)^2 \sigma_e^2 \lp 1 - \gamma_e \sigma_e \rp + (a S_a (s_b-\langle s_b \rangle_{\mathrm{fw}}) + S_\Delta\langle s_b \rangle_{\mathrm{fw}})^2 \mathrm{CV}_{\alpha}^2 \,.
\end{align}

\subsection{Comparison between extrinsic and time-additive noise models for exponentially-growing cells}

In time-additive noise models \cite{amir_cell_2014}, the interdivision time follows
\begin{equation}
	\tau_d = \frac{1}{\alpha}\ln\lp \frac{f(s_b)}{s_b} \rp + \eta_\tau \,,
\end{equation}
where noise $\eta_\tau$ is independent of $s_b$ and $\alpha$. 
In the small noise limit, the extrinsic noise model predicts for the interdivision time:
\begin{equation}
	\tau_d \approx \frac{1}{\alpha}\ln\lp \frac{f(s_b,\alpha)}{s_b} \rp +\frac{\eta_e(\alpha)}{\alpha} \,.
\end{equation}
These two models are thus equivalent in this limit for constant growth rate, with $\eta_\tau \equiv \eta_e(\alpha)/\alpha$. In the presence of fluctuating growth rates however, time-additive noise models assume no size control sensitivity in growth rate through neither $f$ nor $\eta_\tau$. The independence of $\eta_\tau$ from $\alpha$ implies that the corresponding extrinsic noise $\eta_e$ on division size scales with $\alpha$, such that $\sigma_e^2(\alpha) = \alpha^2 \langle \eta_\tau^2 \rangle \approx \langle \alpha \rangle^2 \langle \eta_\tau^2 \rangle (1+ 2 (\alpha-\langle \alpha \rangle)/\langle \alpha \rangle)$ around the mean growth rate. This imposes an extrinsic variance sensitivity $S_\sigma=2$.

In another interpretation of "time-additive" noise, noise is added directly to the log-size \cite{hein_asymptotic_2024}: $\ln(s_d) = \ln(f(s_b)) + \eta_l$. Again, the multiplicative scaling with respect to size is consistent with that of the extrinsic noise model. However, in these models $\eta_l$ is assumed to be independent from single cell growth, leading effectively to $S_\sigma=0$.

\section{Supplementary data figures}

\subsection{Extrinsic noise model explains data across conditions}

\subsubsection{Comparison between fits given by different noise models}
\label{sec_noise_mod_fits}

\begin{figure}
	\centering
	\includegraphics[width=\linewidth]{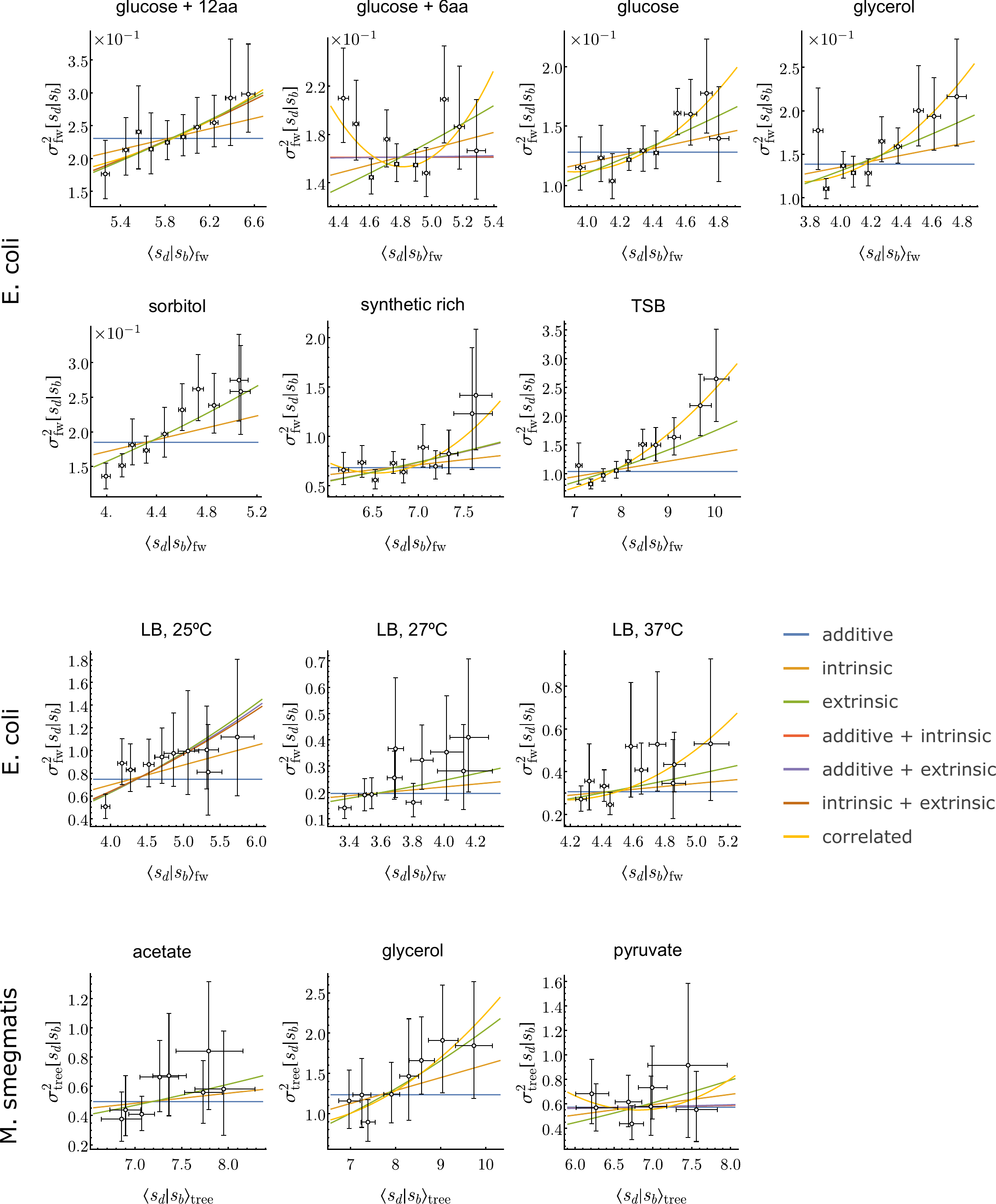}
	\caption{\textbf{Best fits from different noise models across data.} Error bars denote the $95\%$ confidence interval of the estimate determined by bootstrapping. The legend applies to all plots and only non-redundant models, i.e. which give different predictions, are shown. Fit parameters are provided in \cref{tab:fits}.}
	\label{fig:model_fits}
\end{figure}

\renewcommand{\arraystretch}{2}
\begin{table}[]
	\centering
	\hspace*{-1cm}
	\begin{tabular}{|c|c|c|c|c|c|c|c|c|}
		\hline
		& & $\sigma_a^2$ & $\sigma_i^2 x$ & $\sigma_e^2 x^2$ & $\sigma_a^2 + \sigma_i^2 x$ & $\sigma_a^2+ \sigma_e^2 x^2$ & $\sigma_i^2 x + \sigma_e^2 x^2$ & $\sigma_a^2+ (\sigma_i^2 + 2 \langle \eta_a \eta_e \rangle) x + \sigma_e^2 x^2$ \\
		\hline
		\multirow{7}{*}{\rotatebox{90}{E. coli}} & glucose+12aa & \makecell{$0.231$\\$-39.1071$} & \makecell{$0.0396 x$\\$-49.4401$} & \makecell{$0.00673 x^2$\\$-56.5091$} & \makecell{$0.0396 x$\\$-47.2622$}  & \makecell{$0.0136 + 0.00634 x^2$\\$-54.5147$} & \makecell{$0.00444 x + 0.00598 x^2$\\$-54.4984$} & \makecell{$0.520 - 0.174 x + 0.0212 x^2$\\$-52.5024$} \\
		\cline{2-9}
		& glucose+6aa & \makecell{$0.161$\\$-43.8711$} & \makecell{$0.0336 x$\\$-42.8439$} & \makecell{$0.00699 x^2$\\$-40.3669$} & \makecell{$0.160 + 0.000287 x$\\$-41.6933$} & \makecell{$0.157 + 0.000182 x^2$\\$-41.6963$} & \makecell{$0.0336 x$\\$-40.6661$} & \makecell{$5.55 - 2.24 x + 0.233 x^2$\\$-43.2379$} \\
		\cline{2-9}
		& glucose & \makecell{$0.128$\\$-45.5849$} & \makecell{$0.0298 x$\\$-51.1136$} & \makecell{$0.00690 x^2$\\$-56.1722$} & \makecell{$0.0298 x$\\$-48.9358$} & \makecell{$0.00690 x^2$\\$-53.9944$} & \makecell{$0.00690 x^2$\\$-53.9944$} & \makecell{$1.50 - 0.707 x + 0.0900 x^2$\\$-54.3864$} \\
		\cline{2-9}
		& glycerol & \makecell{$0.139$\\$-37.7083$} & \makecell{$0.0338 x$\\$-42.1764$} & \makecell{$0.00819 x^2$\\$-46.547$} & \makecell{$0.0338 x$\\$-39.9986$} & \makecell{$0.00819 x^2$\\$-44.3692$} & \makecell{$0.00819 x^2$\\$-44.3692$} & \makecell{$1.49 - 0.739 x + 0.0996 x^2$\\$-44.7327$} \\
		\cline{2-9}
		& sorbitol & \makecell{$0.185$\\$-29.7314$} & \makecell{$0.0429 x$\\$-36.5896$} & \makecell{$0.00983 x^2$\\$-46.3431$} & \makecell{$0.0429 x$\\$-34.4118$} & \makecell{$0.00983 x^2$\\$-44.1652$} & \makecell{$0.00983 x^2$\\$-44.1652$} & \makecell{$-0.0437 x + 0.0197 x^2$\\$-50.0737$} \\
		\cline{2-9}
		& synthetic rich & \makecell{$0.683$\\$-5.56893$} & \makecell{$0.102 x$\\$-8.11931$} & \makecell{$0.0151 x^2$\\$-9.20157$} & \makecell{$0.102 x$\\$-5.94148$} & \makecell{$0.0208 + 0.0146 x^2$\\$-7.02804$} & \makecell{$0.00176 x + 0.0148 x^2$\\$-7.02409$} & \makecell{$17.4 - 5.12 x + 0.391 x^2$\\$-7.65295$} \\
		\cline{2-9}
		& TSB & \makecell{$1.03$\\$13.6925$} & \makecell{$0.135 x$\\$8.32108$} & \makecell{$0.0174 x^2$\\$0.473941$} & \makecell{$0.135 x$\\$10.4989$} & \makecell{$0.0174 x^2$\\$2.65177$} & \makecell{$0.0174 x^2$\\$2.65177$} & \makecell{$4.05 - 1.20 x + 0.104 x^2$\\$-7.29107$} \\
		\hline
		\multirow{3}{*}{\rotatebox{90}{E. coli}} & LB,\SI{25}{\degree C} & \makecell{$0.747$\\$4.70278$} & \makecell{$0.174 x$\\$-1.67966$} & \makecell{$0.0393 x^2$\\$-4.31141$} & \makecell{$0.174 x$\\$0.498166$} & \makecell{$0.0425 + 0.0372 x^2$\\$-2.18469$} & \makecell{$0.0287 x + 0.0329 x^2$\\$-2.25418$} & \makecell{$0.0287 x + 0.0329 x^2$\\$0.081136$} \\
		\cline{2-9}
		& LB,\SI{27}{\degree C} & \makecell{$0.196$\\$-18.1056$} & \makecell{$0.0552 x$\\$-20.4635$} & \makecell{$0.0154 x^2$\\$-22.6159$} & \makecell{$0.0552 x$\\$-18.2857$} & \makecell{$0.0154 x^2$\\$-20.4381$} & \makecell{$0.0154 x^2$\\$-20.4381$} & \makecell{$-0.0968 x + 0.0421 x^2$\\$-19.8816$} \\
		\cline{2-9}
		& LB,\SI{37}{\degree C} & \makecell{$0.306$\\$-13.5195$} & \makecell{$0.0690 x$\\$-15.5732$} & \makecell{$0.0155 x^2$\\$-17.4638$} & \makecell{$0.0690 x$\\$-13.3954$} & \makecell{$0.0155 x^2$\\$-15.286$} & \makecell{$0.0155 x^2$\\$-15.286$} & \makecell{$6.15 - 2.83 x + 0.339 x^2$\\$-15.1051$} \\
		\hline
		\multirow{3}{*}{\rotatebox{90}{M. smegmatis}} & acetate & \makecell{$0.495$\\$-5.08756$} & \makecell{$0.0691 x$\\$-7.2191$} & \makecell{$0.00960 x^2$\\$-9.14696$} & \makecell{$0.0691 x$\\$-4.9859$} & \makecell{$0.00960 x^2$\\$-6.91375$} & \makecell{$0.00960 x^2$\\$-6.91375$} & \makecell{$-0.114 x + 0.0253 x^2$\\$-5.96649$} \\
		\cline{2-9}
		& glycerol & \makecell{$1.23$\\$9.55788$} & \makecell{$0.161 x$\\$4.36306$} & \makecell{$0.0205 x^2$\\$-0.135274$} & \makecell{$0.161 x$\\$6.59627$} & \makecell{$0.0205 x^2$\\$2.09793$} & \makecell{$0.0205 x^2$\\$2.09793$} & \makecell{$2.75 - 0.714 x + 0.0663 x^2$\\$4.18929$} \\
		\cline{2-9}
		& pyruvate & \makecell{$0.572$\\$-7.10273$} & \makecell{$0.0846 x$\\$-6.63541$} & \makecell{$0.0124 x^2$\\$-5.17032$} & \makecell{$0.499 + 0.0109 x$\\$-4.88019$} & \makecell{$0.525 + 0.00103 x^2$\\$-4.88768$} & \makecell{$0.0846 x$\\$-4.4022$} & \makecell{$8.82 - 2.43 x + 0.179 x^2$\\$-3.19031$} \\
		\hline
	\end{tabular}
	\caption{\textbf{Best fit parameters and AIC values for all conditions (rows) and all noise models (columns).} In each cell, the first line shows the noise models $\sigma^2[s_d|s_b] = g(\langle s_d | s_b \rangle)$, with $x=\langle s_d | s_b \rangle$, and the second line shows the AIC.}
	\label{tab:fits}
\end{table}

In this section, we show how the different noise models fit the data. We show in \cref{fig:model_fits} the best fits for all non-redundant models, i.e. excluding models giving the same fit as explained below, and we give in \cref{tab:fits} the values of the best fit parameters and the AIC values. 
The seven models are defined by $\sigma^2[s_d|s_b] = g(\langle s_d | s_b \rangle)$, with $g(x)= \sigma_a^2$ for the additive noise model, $g(x)= \sigma_i^2 x$ for the intrinsic noise model and $g(x)= \sigma_e^2 x^2$ for the extrinsic noise model, and by combinations of these functions. 

The last model we consider is a combination of the three noises where we allow additive and extrinsic noises to be correlated. In this case, the total variance reads 
\begin{equation}
	\sigma_\mathrm{fw}^2[s_d|s_b] = \sigma_a^2 + \lp \sigma_i^2 + 2 \langle \eta_a \eta_e \rangle_\mathrm{fw} \rp \langle s_d | s_b \rangle + \sigma_e^2 \langle s_d | s_b \rangle^2 \,,
\end{equation}
where additive-extrinsic correlations lead to an intrinsic-like contribution to the noise which can be negative. For this model we therefore do not require that the coefficient linear in $\langle s_d | s_b \rangle$ be positive.
Moreover, since a negative intrinsic contribution can only arise from negatively correlated additive and extrinsic noises, we exclude the fits produced by the correlated model if the linear coefficient is negative while there is no additive noise, which is the case for \textit{E. coli} in sorbitol and LB medium at \SI{27}{\degree C}, and for \textit{M. smegmatis} in acetate. The only condition which is best described by a correlated noise model is TSB. 

In \cref{tab:fits}, each cell shows the best fit on the first line and the value of the AIC on the second line. In many cases, combined-noise models give the same fits as noise models with less parameters, like the additive+extrinsic and the extrinsic noise models for \textit{E. coli} in glucose for example. In these cases, we call redundant the model with the largest number of parameters, and we do not show it in \cref{fig:model_fits}. Note that the AIC of a redundant model is always larger than that of the corresponding simpler model, because the goodness of fit is the same, but the penalty for the number of parameters is larger. 
The probability that a model $m$ describes the data, shown in the main text in 
fig. 3g-i,
is computed with the AIC as $p(m)=e^{-\mathrm{AIC}(m)/2}/\sum_{m' \in \mathrm{models}} e^{-\mathrm{AIC}(m')/2}$ where the sum runs through non-redundant models only.

\subsubsection{Added size distributions collapse is incompatible with the extrinsic noise model}

\begin{figure}
	\centering
	\includegraphics[width=\linewidth]{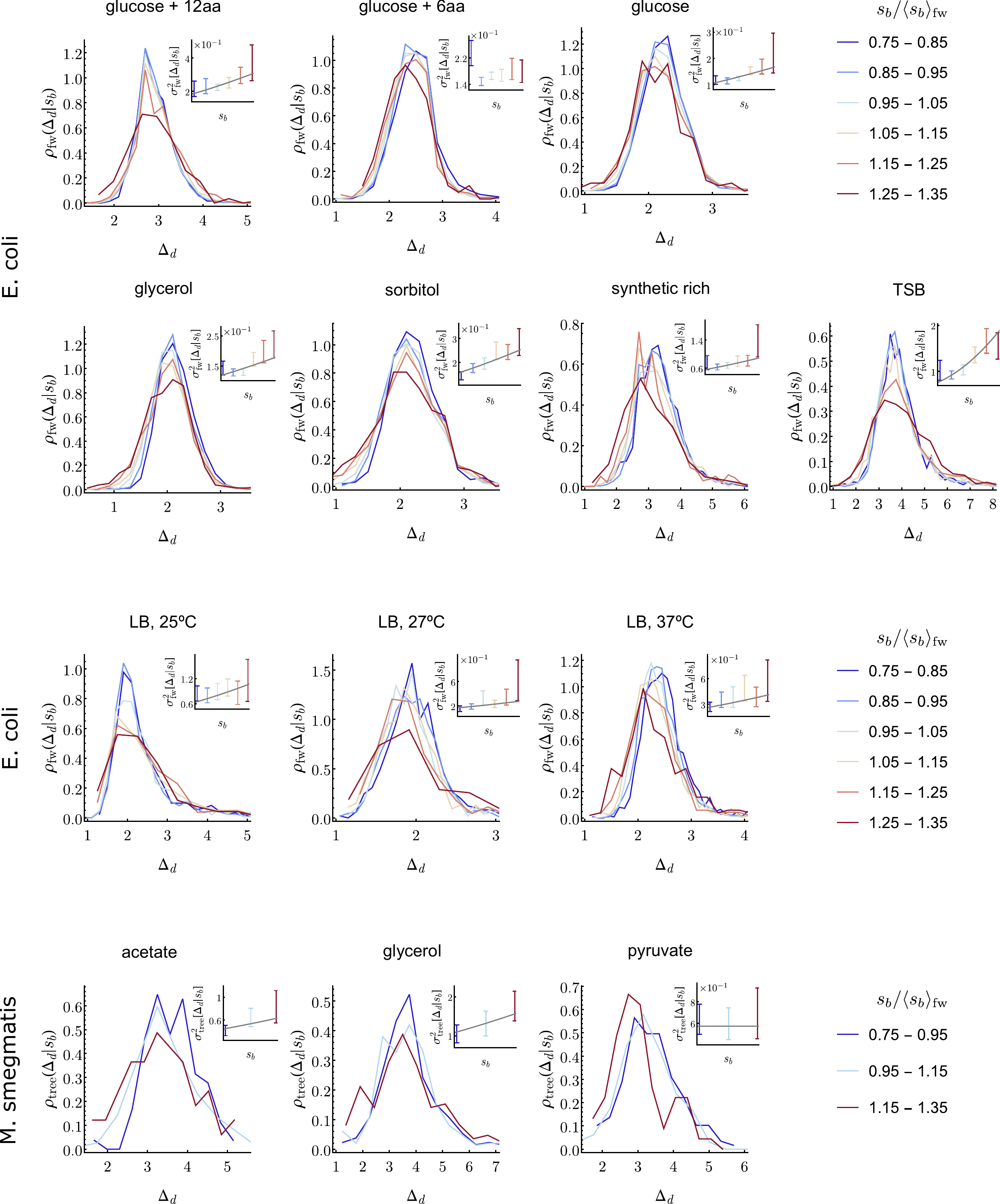}
	\caption{\textbf{Dependence of the added size distributions on birth size across conditions.} Insets show the variance of the conditional added size distribution against birth size. Error bars denote the $95\%$ confidence interval of the estimate determined by bootstrapping. The grey curves indicate predictions from the best noise model determined above.}
	\label{fig:Delta_distrib}
\end{figure}

In \cref{fig:Delta_distrib}, we reproduce the plots of the added size distributions for different birth sizes for \textit{E. coli} in different growth media from \cite{taheri-araghi_cell-size_2015}. Unlike in 
fig. 3d
in the main text where analysed birth sizes span the interval from the $2.5 \%$ quantile to the $97.5 \%$ quantile, resulting in $s_b \in [0.77,1.34] \langle s_b \rangle_\mathrm{fw}$, here we analyse data with birth sizes in the interval $s_b \in [0.75,1.35] \langle s_b \rangle_\mathrm{fw}$ to ease the comparison with the analysis from \cite{taheri-araghi_cell-size_2015}.

Like for sorbitol in the main text, we observe that the distributions do not collapse. We report in insets the added size variance as a function of the birth size, and show that they are consistent with the predictions of the extrinsic noise model, shown in grey, with fitted parameters from the analysis described in the previous section.
For TSB, the best fit from the correlated noise model is shown instead, and no prediction is shown for glucose+6aa which exhibits no clear trend.

We perform the same analysis for \textit{E. coli} data at different temperatures from \cite{tanouchi_noisy_2015} and for \textit{M. smegmatis} data in different growth media from \cite{priestman_mycobacteria_2017}, and show the results in \cref{fig:Delta_distrib}. Similarly, the distributions do not collapse and the agreement between the increase of added size variance with birth size and the prediction of the extrinsic noise model is shown in insets. For \textit{M. smegmatis} in pyruvate, the inset instead shows the predictions from the additive noise model.

\subsubsection{Ignoring extrinsic noise leads to significant error}

We explore here the error made on the standard deviation of added size when modelling noise as additive when it is actually extrinsic in the data, and compare it to the error made on the mean added size when modelling the mechanism of cell size control as an adder when $a$ is close to but actually different from $1$ in the data. 

In data the mean added size is rarely perfectly independent of the birth size, yet slopes $a$ close to $1$ are generally called near-adder and modelled as adders with $a=1$. 
This simplification leads to a relative error 
\begin{equation}
	\epsilon_\mu = \frac{\langle \Delta_d | s_b^\mathrm{max} \rangle}{ \langle \Delta_d | s_b^\mathrm{min} \rangle} -1
\end{equation}
on the mean added size between the smallest and the biggest newborn cells, with $\langle \Delta_d|s_b \rangle = (a-1) s_b +b$. This relative error is null when the data follow an adder, and otherwise scales as $\epsilon_\mu\approx (a-1)(s_b^\mathrm{max} - s_b^\mathrm{min})/\langle s_b \rangle $ close to the adder (first order term in the expansion with small parameter $a-1$).

Similarly, assuming additive noise when it is actually extrinsic leads to the relative error 
\begin{equation}
	\epsilon_\sigma = \frac{\sigma[\Delta_d | s_b^\mathrm{max}]}{ \sigma[\Delta_d | s_b^\mathrm{min}]} -1
\end{equation}
on the standard deviation, with $\sigma[\Delta_d | s_b]=\sigma_e (a s_b +b)$. This relative error is null if noise is purely additive, and otherwise scales as $\epsilon_\sigma \approx (s_b^\mathrm{max} - s_b^\mathrm{min})/(\langle s_b \rangle + s_b^\mathrm{min}) + 2(a-1)\langle s_b \rangle(s_b^\mathrm{max} - s_b^\mathrm{min})/(\langle s_b \rangle + s_b^\mathrm{min})^2$ close to the adder.

Close to the adder, there is thus an affine relationship between these two relatives errors, given by:
\begin{equation}
	\label{eq_err_ext_lin}
	\epsilon_\sigma \approx \frac{s_b^\mathrm{max} - s_b^\mathrm{min}}{\langle s_b \rangle + s_b^\mathrm{min}} + \frac{2\langle s_b \rangle^2}{(\langle s_b \rangle + s_b^\mathrm{min})^2} \epsilon_\mu \,.
\end{equation}
For small $\epsilon_\mu$ the first term typically dominates and makes $\epsilon_\sigma$ larger than $|\epsilon_\mu|$.

\begin{figure}
	\centering
	\includegraphics[width=0.8\linewidth]{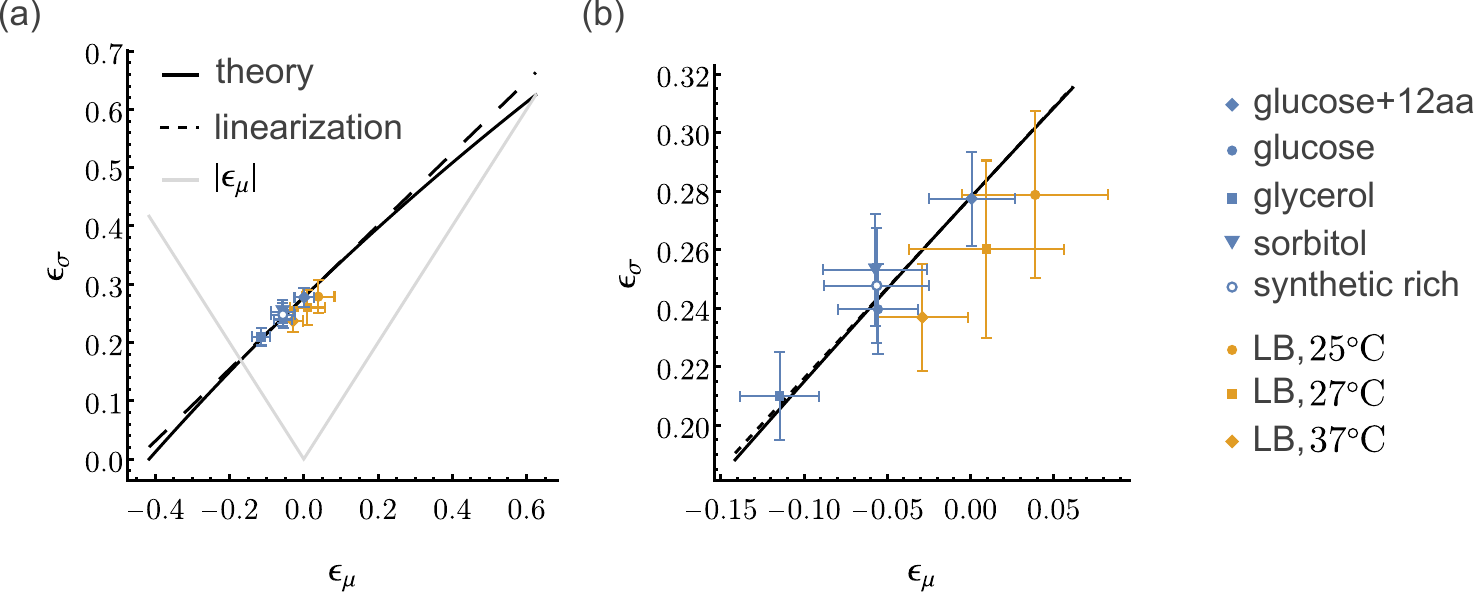}
	\caption{Relative error on the standard deviation of added size when assuming that noise is additive instead of extrinsic against relative error on the mean added size when assuming that $a=1$.}
	\label{fig:error_ext}
\end{figure}

We computed these relative errors in data for \textit{E. coli} in different growth media and temperatures, for conditions which follow the extrinsic noise model (thus excluding glucose+6aa and TSB). 
We used $s_b^\mathrm{min}=0.8\langle s_b\rangle$ and $s_b^\mathrm{max}=1.3\langle s_b\rangle$, corresponding to the middle values of the bins with extremal sizes in the analysis of \cref{fig:Delta_distrib} (respectively blue and red). The added size means and standard deviations for these birth sizes were computed using the linear map and the extrinsic noise definition with fitted $a$, $b$ and $\sigma_e$.
Results are shown in \cref{fig:error_ext}a, where the plain black line is a parametric plot of $\epsilon_\sigma$ against $\epsilon_\mu$ when varying $a$ from $0$ to $2$, the dashed line indicates the linear regime close to the adder given by \cref{eq_err_ext_lin}, and the grey cone shows where $\epsilon_\sigma \geq |\epsilon_\mu|$. A zoomed in version focusing on the data is also proposed in \cref{fig:error_ext}b. 
We observe that in all conditions $\epsilon_\sigma > |\epsilon_\mu |$ where $\epsilon_\mu$ ranges from $-12\%$ (in glycerol where $a \approx 0.77$) to $4\%$ (in LB at \SI{25}{\degree C} where $a \approx 1.09$) and $\epsilon_\sigma$ ranges from $21\%$ (in glycerol) to $28\%$ (in LB at \SI{25}{\degree C}).

We therefore conclude that, even in conditions where assuming an adder for the mean added size has negligible consequences, failing to account for extrinsic noise fluctuations would lead to significant errors on the standard deviation of added size.

\subsection{Cell size control depends on single-cell growth across conditions}

\subsubsection{Sensitivity of the linear map and of the noise model on single-cell growth}

\begin{figure}
	\centering
	\includegraphics[width=\linewidth]{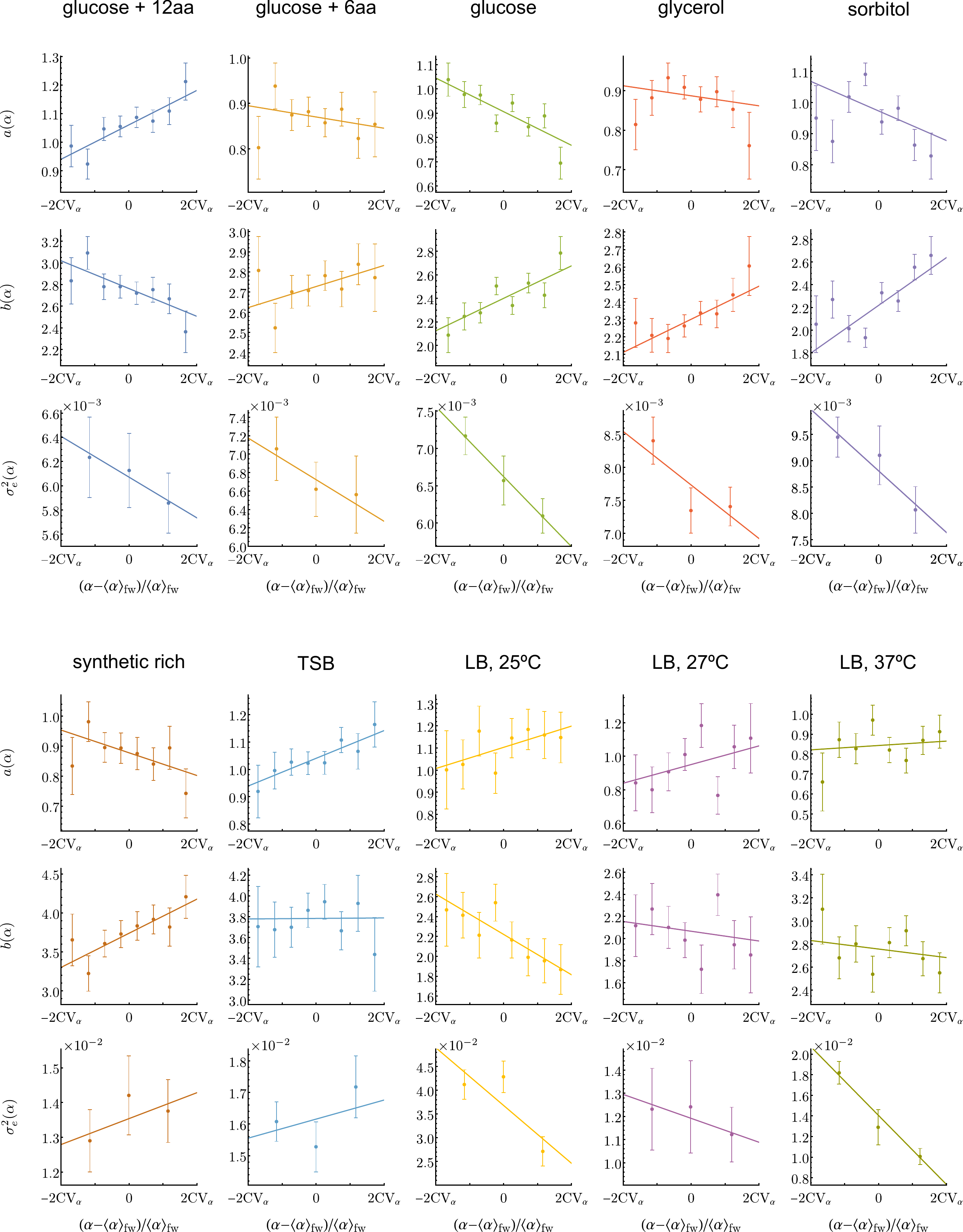}
	\caption{\textbf{Dependence of size control on single-cell growth across conditions.} Slope $a(\alpha)$, intercept $b(\alpha)$ and extrinsic variance $\sigma_e^2(\alpha)$ against single-cell growth rates $\alpha$, where the error bars indicate $\pm$ the standard error on the fitted parameters.}
	\label{fig:linmap_vs_alpha}
\end{figure}

In this section, we show how size control depends on single-cell growth rates for \textit{E. coli} across conditions. 

For each condition, single cell growth rates $\alpha$ between the $2.5\%$ quantile and the $97.5\%$ quantile were divided into 8 bins. 
In each bin, the slope $a(\alpha)$ and the intercept $b(\alpha)$ of the linear map were obtained with a linear fit of division sizes against birth sizes. In \cref{fig:linmap_vs_alpha}, we show the dependencies of the slope and intercept on single-cell growth rates, where error bars are given by the standard error of the fit parameters.

\begin{figure}
	\centering
	\includegraphics[width=\linewidth]{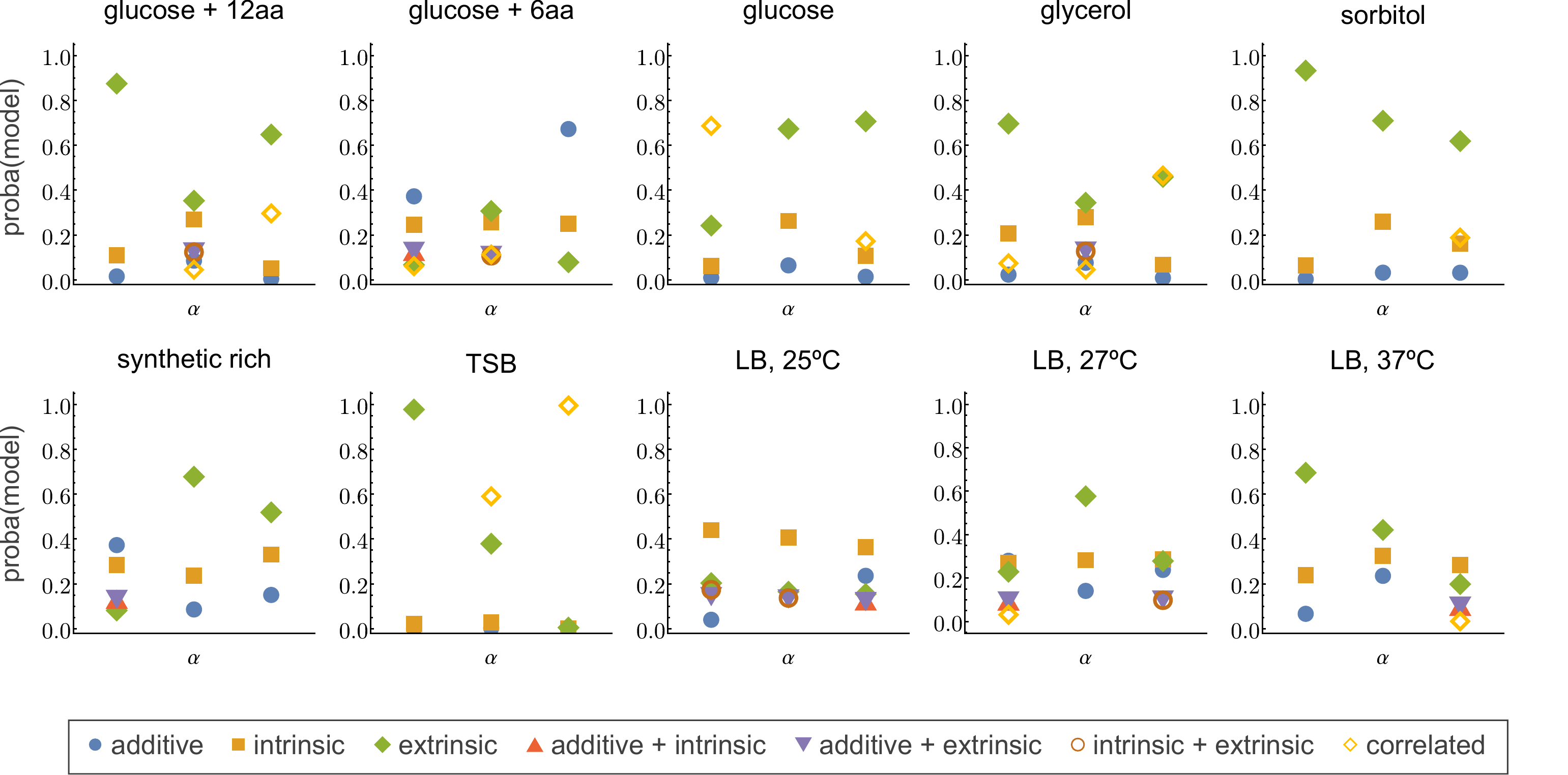}
	\caption{\textbf{Probabilities of different noise models across conditions and single-cell growth rates.} For each condition single-cell growth rates were divided in three bins, and model probabilities were computed as proba$(m)=e^{-\mathrm{AIC}(m)/2}/\sum_{m' \in \mathrm{models}} e^{-\mathrm{AIC}(m')/2}$. Only non-redundant models are shown and included in the sum.}
	\label{fig:noise_mod_pb_alpha}
\end{figure}

To quantify the noise terms, we binned the growth rates in the same interval into 3 bins (the bin number is reduced from 8 to 3 because the noise analysis requires a second binning on birth sizes next), and we repeated the analysis of \cref{sec_noise_mod_fits} for each bin.  
Note that we excluded correlated noise models producing unrealistic additive noise amplitudes $\sigma_a>\langle s_b \rangle$ (which applied to synthetic rich).
We show in \cref{fig:noise_mod_pb_alpha} the model probabilities for all conditions and all single-cell growth rates.
We conclude that, except for glucose+6aa and TSB which do not follow an extrinsic noise model when fluctuations in $\alpha$ are ignored, the extrinsic noise model explains division size fluctuations across a majority of conditions and growth rates. Indeed, for all conditions extrinsic noise is the best model for at least two out of three growth rate bins. The only notable exception is the condition LB at \SI{27}{\degree C}, for which the best noise model is intrinsic for all growth rates and yet the best overall fit has been shown to be extrinsic in \cref{sec_noise_mod_fits}. However this condition is noisy and as a result the model probabilities are all below $50\%$, including the best model. For this reason, we consider this condition to be extrinsic as well based on a majority-voting argument from all the other conditions.

We show the values of extrinsic noise variances against growth rates across conditions in \cref{fig:linmap_vs_alpha}. Note that extrinsic variances for glucose+6aa and TSB are shown for completeness, but the extrinsic noise is not the best model for these conditions.

\subsubsection{Linear map sensitivity on single-cell growth agrees with direct measure from added size}

In this section, we show that the added size sensitivity on growth rate predicted from the sensitivities of linear map slope and intercept in eq. 5 in the main text is in excellent agreement with a direct analysis of added size versus growth rate across conditions. 

\begin{figure}
	\centering
	\includegraphics[width=\linewidth]{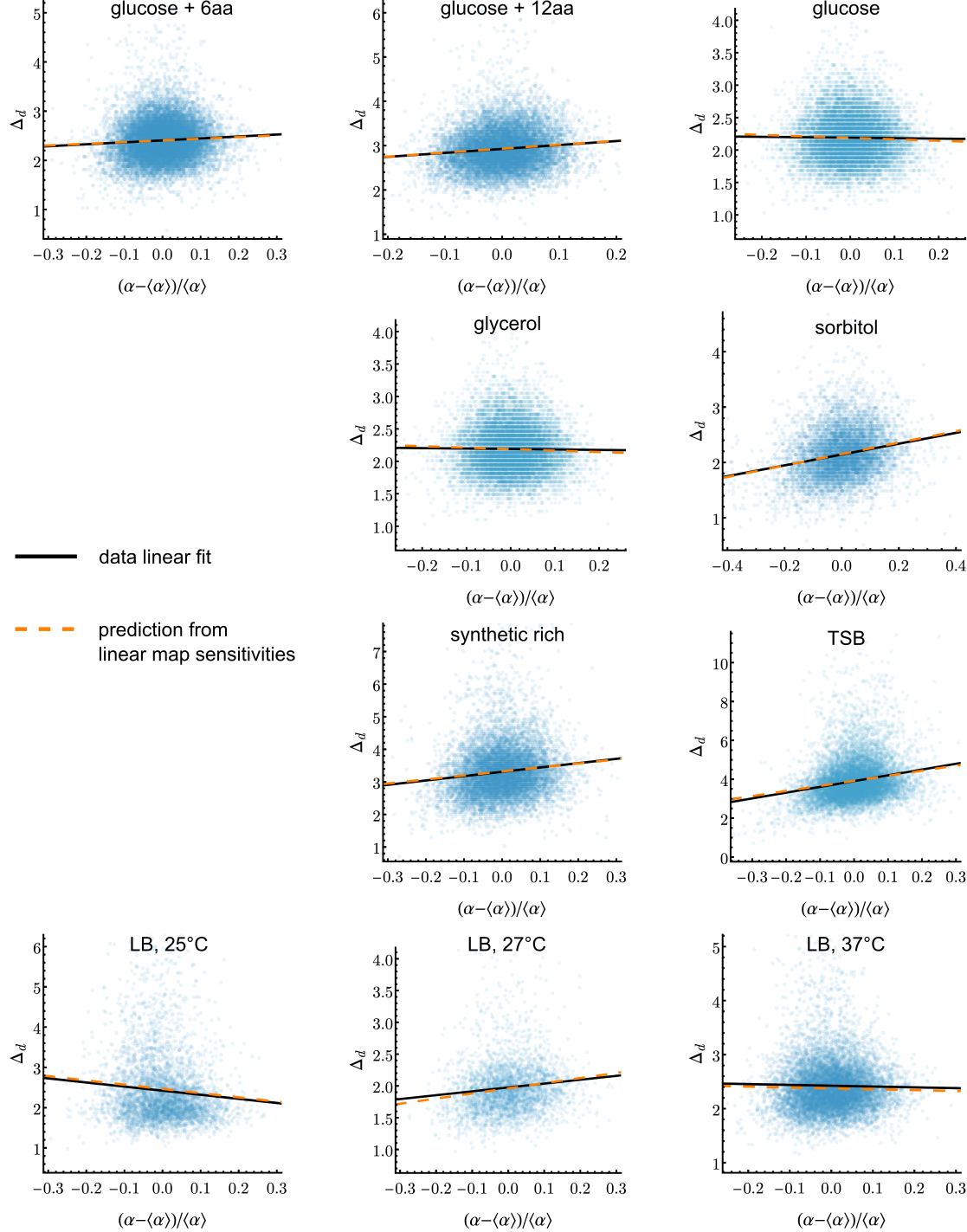}
	\caption{\textbf{Added size against single-cell growth rate across conditions.} Direct linear fits of scatter data (black) are in excellent agreement with predictions from fitted linear map sensitivities.}
	\label{fig:Delta_vs_alpha}
\end{figure}

In \cref{fig:Delta_vs_alpha}, we show scatter plots of added sizes against single-cell growth rates across conditions. The black lines represent the best linear fit of the data, and the orange dashed lines represent the predictions from the linear map sensitivities: $\langle \Delta_d | \alpha \rangle = b/(2-a) (1+ (a S_a + (2-a) S_b) (\alpha-\langle \alpha \rangle)/\langle \alpha \rangle)$. 

\subsection{Noise minimisation is in good agreement with general extrinsic noise model}

\begin{figure}
	\centering
	\includegraphics[width=\linewidth]{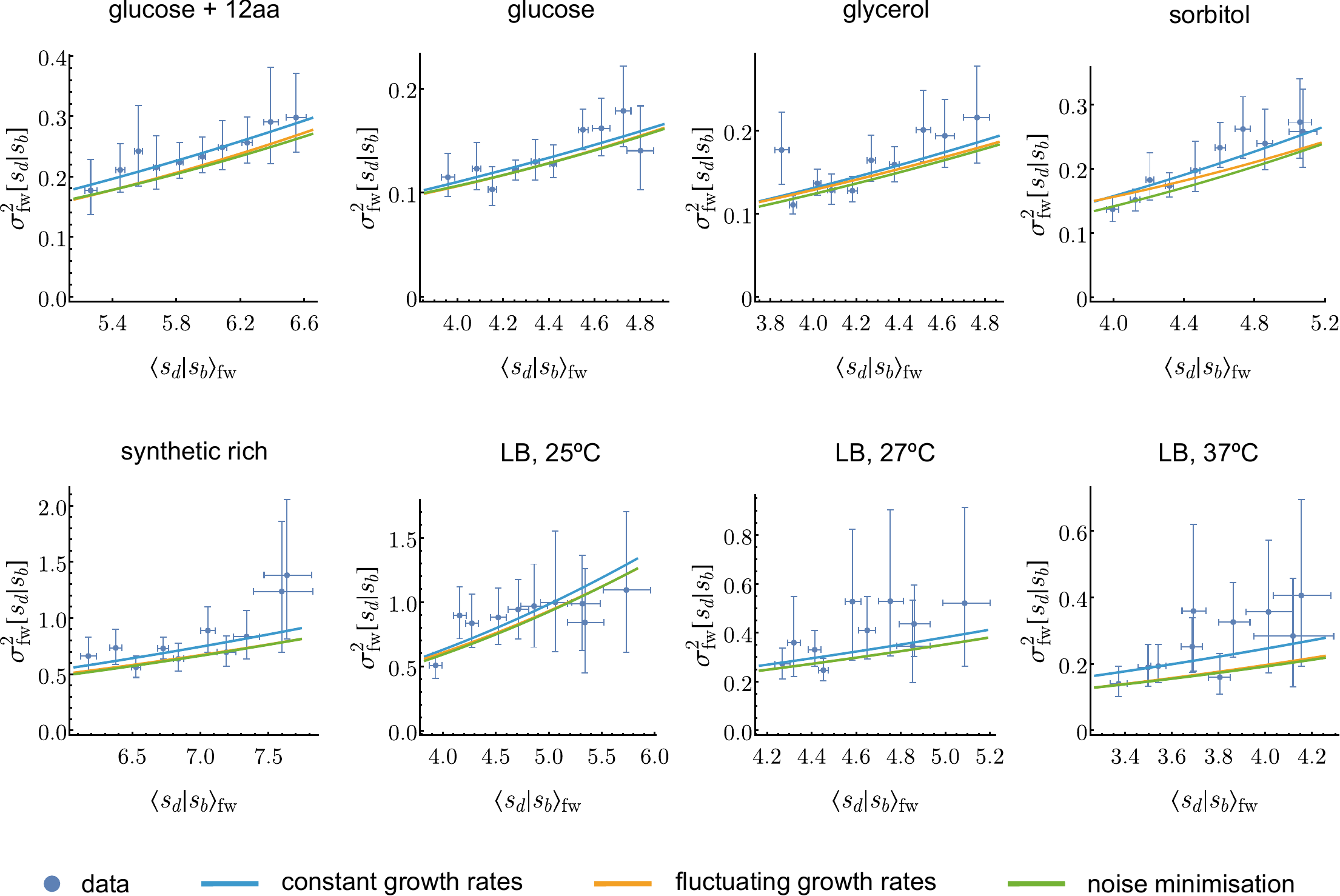}
	\caption{The extrinsic noise model fitted when ignoring single-cell growth rate fluctuations (blue), the noise model obtained in eq. 9 in the main text by coarse-graining fluctuating growth rates (orange), and the model of noise minimisation obtained by setting $S_\Delta=0$ in eq. 9 (green) are all in agreement with the data.}
	\label{fig:fit_cond_var}
\end{figure}

In \cref{fig:fit_cond_var}, we show the conditional variance of division size against the conditional mean division size across conditions following an extrinsic noise model, and the agreement with three noise models. This analysis complements the one shown in fig. 6c in the main text for \textit{E. coli} in glucose. 
The best extrinsic fits when ignoring growth rate fluctuations, obtained in \cref{sec_noise_mod_fits}, are shown in blue; the noise models with additive, intrinsic and extrinsic contributions coming from the coarse-graining over fluctuating growth rates, derived in eq. 9 in the main text, are shown in orange; and the predictions obtained when minimising growth-rate-induced noise by setting $S_\Delta=0$ in eq. 9 are shown in green. 
The consistency of these three noise models with the data indicates that growth-rate-induced noise is minimised in \textit{E. coli}.

\subsection{Noise analysis for \textit{E. coli} in slow growth condition}

\begin{figure}
	\centering
	\includegraphics[width=\linewidth]{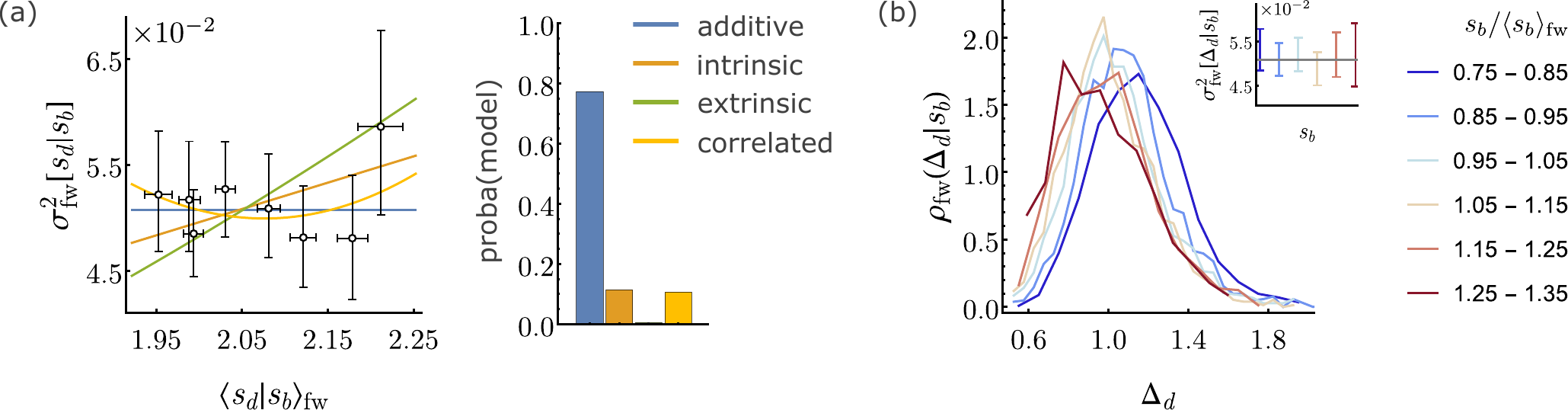}
	\caption{\textbf{Additive noise model captures \textit{E. coli} behaviour in slow growth condition M9+glycerol.} Model probabilities are computed based on the AIC as explained in the caption of \cref{fig:noise_mod_pb_alpha}.}
	\label{fig:slow_growth}
\end{figure}

In this section, we analyse data of \textit{E. coli} in M9+glycerol (slow growth condition) from \cite{nieto_mechanisms_2024}. 
Birth sizes in the interquantile interval $[5\%,95\%]$ were divided into eight bins for the noise model analysis.
Our analysis based on the AIC score reveals that the additive noise model is the most likely to explain the data, as shown in \cref{fig:slow_growth}a (only non-redundant models are shown).
We also show how distributions of added size depend on birth size in \cref{fig:slow_growth}b. For this analysis, like for the other data, birth sizes in the interval $[0.75,1.35] \langle s_b \rangle_\mathrm{fw}$ were divided into six bins. We note a discrepancy between the value $\langle s_b \rangle_\mathrm{fw} \approx 1.0$ we computed and the value $\langle s_b \rangle_\mathrm{fw} \approx 2$ reported in \cite{nieto_mechanisms_2024}.
The shift of distributions towards small added sizes as birth size increases is consistent with the fact that in this slow growth condition, \textit{E. coli} deviates from the adder towards a more sizer-like behaviour (our linear map fit gives $a=0.62$). In inset we show however that the variance of these distributions is independent of birth size, and consistent with the additive noise variance inferred above.

%